\documentclass[12pt,a4paper]{article}
 \usepackage[skins,theorems]{tcolorbox}
\tcbset{highlight math style={enhanced,
  colframe=red,colback=white,arc=0pt,boxrule=1pt}}
  \usepackage{tikz}
  \usepackage{tikz-3dplot}
 \usetikzlibrary{calc}
 \usetikzlibrary{decorations} %
 \usepackage[UKenglish]{babel}
 \usepackage[toc,page]{appendix}
 \usepackage{amsmath}
 \usepackage{amssymb}
 \usepackage{graphicx}
 \usepackage{hhline}
 \usepackage[bf]{caption}
\usepackage{cite}
\usepackage[vcentermath]{youngtab}
\usepackage{geometry}
\usepackage{slashed}
\usepackage{color}
\usepackage{stackrel}
\usepackage{tikz-cd} 
\usepackage{cancel} 
\usepackage{multirow} 
\usepackage{longtable}
\usepackage{ytableau}

\usepackage{tikz}
\usetikzlibrary{shadows}
\usepackage[framemethod=tikz]{mdframed}

\allowdisplaybreaks

\global\mdfdefinestyle{myboxstyle}{%
  shadow=true,
  linecolor=black,
  shadowcolor=black,
  shadowsize=6pt,
  nobreak=false,
  innertopmargin=10pt,
  innerbottommargin=10pt,
  leftmargin=5pt,
  rightmargin=5pt,
  needspace=1cm,
  skipabove=10pt,
  skipbelow=15pt,
  middlelinewidth=1pt,
  afterlastframe={\vspace{5pt}},
  aftersingleframe={\vspace{5pt}},
  tikzsetting={%
draw=black,
very thick} }

\newmdenv[style=myboxstyle]{whitebox} \newmdenv[style=myboxstyle,backgroundcolor=black!20]{graybox}

\definecolor{bluish}{rgb}{0.0, 0.2, 0.4}
\definecolor{darkred}{rgb}{0.7, 0.0, 0.0}
\definecolor{darkgreen}{rgb}{0.0, 0.5, 0.0}

\newmdenv[style=myboxstyle,nobreak=true]{blockwhitebox}
\newmdenv[style=myboxstyle,backgroundcolor=black!20,nobreak=true]{blockgraybox}
\newmdenv[nobreak=true,hidealllines=true]{blockbox}

\usepackage{jheppub}

\newcommand{\bqa}{\begin{eqnarray}}
\newcommand{\eqa}{\end{eqnarray}}

\def\et24{\eta^{24}}
\def\oet24{\frac1{\eta^{24}}}

\hypersetup{
    pdftitle={},
    pdfauthor={},
    pdfsubject={}
}
\numberwithin{equation}{section}
\numberwithin{table}{section}

\makeatletter

\newcommand{\be}{\begin{equation}}
\newcommand{\ee}{\end{equation}}
\newcommand{\beq}{\begin{equation}}
\newcommand{\eeq}{\end{equation}}
\newcommand{\ba}{\begin{aligned}}
\newcommand{\ea}{\end{aligned}}

\newcommand{\bea}{\begin{eqnarray}}
\newcommand{\eea}{\end{eqnarray}}

\newcommand\bi{\begin{itemize}}
\newcommand\ei{\end{itemize}}

\def\Im{\mathop{\mathrm{Im}}\nolimits}

\def\Tr{\mathop{\mathrm{Tr}}\nolimits}

\def\unit{{1\kern-.65ex {\rm l}}}
\def\1{{1\kern-.65ex {\rm l}}}

\def\QQ{\mathcal{Q}_{\mathbf{ALE}_n,\boldsymbol{\kappa}}}
\def\QQTN{\mathcal{Q}_{\mathbf{TN}_n,\boldsymbol{\kappa}}}

\makeatother

\begin{document}

\title{The ALE Partition Functions of M-Strings}

\author[\sharp\dagger]{Michele Del Zotto}
\author[\sharp\star\natural]{and Guglielmo Lockhart}

\affiliation[\sharp]{Department of Mathematics, Uppsala University, 75106, Uppsala, Sweden}
\affiliation[\dagger]{Department of Physics and Astronomy, Uppsala University, 75120, Uppsala, Sweden}
\affiliation[\star]{Theory Department, CERN,
 Geneva 23, CH-1211, Switzerland}
\affiliation[\natural]{Bethe Center for Theoretical Physics, Universit\"at Bonn, D-53115, Germany}

\emailAdd{michele.delzotto@math.uu.se}
\emailAdd{glockhar@uni-bonn.de}

\abstract{We compute the equivariant partition function of the six-dimensional M-string SCFTs on a background with the topology of a product of a two-dimensional torus and an ALE singularity. We determine the result by exploiting BPS strings probing the singularity, whose worldvolume theories we determine via a chain of string dualities. A distinguished feature we observe is that for this class of background the BPS strings' worldsheet theories become relative field theories that are sensitive to finer discrete data generalizing to 6d the familiar choices of flat connections at infinity for instantons on ALE spaces. We test our proposal against a conjectural 6d $\mathcal{N}=(1,0)$ generalization of the Nekrasov master formula, as well as against known results on ALE partition functions in four dimensions.
}

\maketitle

\section{Introduction and outlook}
Six dimensional $\mathcal N=(2,0)$ supersymmetric conformal field theories are systems at the verge of inconsistency whose existence was derived long ago within string theory \cite{Witten:1995zh,Strominger:1995ac}. 
One of the biggest challenges in studying these theories is that they cannot have a conventional Lagrangian description \cite{Chang:2018xmx}. This is among the reasons for which at present we know of few observables that can be exactly computed for 6d SCFTs. Important examples include the partition functions or large $N$ free energies of these systems on backgrounds such as $T^2\times S^4$ and $S^1 \times S^5$ as well as on the $\Omega$-background $(T^2 \times \mathbb C^2)_{\epsilon_1,\epsilon_2}$, possibly in presence of defects --- see \emph{e.g.}
\cite{Kim:2012ava,Kallen:2012zn,Kim:2012tr,Fukuda:2012jr,Lockhart:2012vp,Kim:2012qf,Minahan:2013mtr,Yagi:2013fda,Cordova:2013cea,Haghighat:2013gba,Kim:2016usy,Hayling:2017cva,Agarwal:2018tso,Nahmgoong:2019hko,Lee:2020rns}. The main purpose of this paper is to study a different observable for the M-string SCFTs \cite{Haghighat:2013gba}, the so called (equivariant) \textit{ALE partition function}, obtained coupling the SCFT to an $\Omega$-deformed background geometry of the form $(T^2 \times \mathbb C^2 /\mathbb Z_n)_{\epsilon_1,\epsilon_2}$, with a metric which is the product of the standard metric on $T^2$ and the ALE metric of $A_{n-1}$ type on $\mathbb C^2 /\mathbb Z_n$.

There are several motivations for our study. First of all, M-strings give an indirect way of capturing features of $(2,0)$ theories, that can be exploited to probe certain conjectural properties of these theories. In this paper in particular we will see that indeed the ALE partition function of M-strings agrees with the 5d uplift of the equivariant ALE partition function of a 4d $\mathcal N=2^*$ pure super Yang-Mills theory of type $U(r)$ \cite{Nekrasov:2003vi,Gasparim:2009sns,Bonelli:2011jx,Bonelli:2011kv}, which gives a very non-trivial consistency check for several claims in the literature about these systems \cite{Douglas:2010iu}. Secondly, the $\Omega$-deformed flat background, the $T^2 \times S^4$ partition function, and the $S^1 \times S^5$ superconformal index only capture information on the local operators of these theories, while the ALE partition functions we consider in this paper are capable of probing finer aspects of their global structure, in particular their relative (chiral) nature \cite{Witten:2009at,Freed:2012bs}.\footnote{\ For the \textit{cognoscenti}: the backgrounds we consider have non-trivial intersecting 3-cycles which give rise to a non-trivial pairing for the Heisenberg algebra of the 6d SCFT with values in the 2-form sector of the defect group \cite{DelZotto:2015isa}. However, our background contains a $T^2$ factor and therefore the possible conformal blocks of the 6d SCFTs at hand can be organized in terms of partition functions of absolute (as opposed to relative) 4d KK theories, corresponding to the discussion in \cite{Tachikawa:2013hya} but keeping the $T^2$ volume finite -- see also the discussion in \cite{Gukov:2020btk,Bashmakov:2022jtl,Bashmakov:2022uek} as well as the talk \href{https://youtu.be/_vnZSoFgdLc?si=W0-fPTzjwwSGaEjg}{\textit{9 vs 10}} by Sergei Gukov at the \textit{Categorical Aspects of Symmetry Conference} held at NORDITA in August 2023, few weeks prior to the publication of this paper. In this paper we focus on the choice that from the rank $r$ M-string SCFT realizes an absolute 4d KK theory with lowest Landau level corresponding to the $\mathcal N=2^*$ $U(r)$ SYM theory.} In fact, as we will see in details below, our results depend on a slightly more refined collection of 6d quantum numbers which give the six-dimensional counterpart of the choice of flat connection at infinity in gauge theory. Thirdly, from the perspective of mathematical physics, the equivariant higher rank Donaldson-Thomas (DT) theory of Nekrasov and Okounkov \cite{Nekrasov:2014nea} was recently related via a stringy correspondence to equivariant partition functions of 5d theories on Taub-NUT and ALE spaces \cite{DelZotto:2021gzy}. From that perspective, the result of this paper can be viewed as a computation of the equivariant higher rank DT theory for the ellipically fibered Calabi-Yau threefold singularities which realize the M-string SCFT in geometric engineering \cite{Heckman:2013pva,Haghighat:2013gba}, via the M-theory/F-theory duality \cite{Vafa:1996xn}.

Since for the ALE partition function the background is non-compact, we can always prescribe boundary conditions at infinity for the scalars in the tensor multiplets of the 6d SCFT. This realizes the SCFTs' maximally abelian phase by Higgsing them away from conformality. This phase is analogous to a Maxwell phase for a non-abelian gauge theory, but in 6d it is governed by a collection of $r$ self-dual tensor fields, whose vevs parametrize the so-called `tensor branch' of the moduli space of the SCFT. The ALE partition functions depend explicitly on the values of these scalar vevs. As is expected from previous results in the case of 4d or 5d theories in the Coulomb branch \cite{Lossev:1997bz,Losev:1997tp,Nekrasov:2002qd,Nekrasov:2003rj}, deforming to this abelian phase makes the computation of supersymmetric partition functions tractable, including on the present background $T^2\times \mathbb{C}^2/\mathbb{Z}_n$. A hallmark of the non-trivial 6d dynamics in such an abelian phase is given by the presence of solitonic BPS strings coupled to the abelian self-dual tensors, which become tensionless at the conformal point. These solitonic BPS strings have universal properties in flat space that can be exploited to unravel deep features of the 6d SCFTs -- see eg. \cite{DelZotto:2018tcj} for a review. One of the main results of this paper is to characterize the structure of the solitonic BPS strings for the M-string SCFTs on an ALE background. Thanks to this characterization, similarly to what happens for the $\Omega$-deformed partition function in flat space, we find that the ALE partition function is captured, neglecting a classical prefactor, by two main contributions: a universal multiplicative factor encoding contributions from BPS particles, and a sum over ALE solitonic BPS strings, each contributing to the partition function via its elliptic genus, that we explicitly compute --- see section \ref{sec:Mtogether} for the precise expressions. The main tool we use to capture the relevant elliptic genera is a chain of string dualities which brings us to a grid of intersecting fivebranes in Type IIB. The BPS strings arise from D3 branes suspended on plaquettes of the grid, which give rise to two-dimensional $\mathcal{N}=(0,4)$ supersymmetric quantum field theories. These 2d QFTs receive contributions from chiral fermions that arise from the fivebrane intersections but in fact exhibit very non-trivial behavior, including being delocalized in nature \cite{Itzhaki:2005tu}.

We present two consistency checks for our proposal. One one hand, we formulate a conjectural 6d $\mathcal{N}=(1,0)$ version of the Nekrasov master formula \cite{Nekrasov:2003vi}, which relates the partition function $T^2\times\widehat{\mathbb{C}^2/\mathbb{Z}_n}$ (where $\widehat{X}$ denotes the blow-up of $X$ at the origin)  to a product of $n$ copies of the partition function on $T^2\times\mathbb{C}^2$. This elliptic version of the Nekrasov master formula can be viewed as a generalization of the elliptic version of the Nakajima-Yoshioka blow-up equations \cite{Nakajima:2003pg,Nakajima:2005fg} for $T^2\times\widehat{\mathbb{C}^2}$ \cite{Gu:2018gmy,Gu:2019dan,Gu:2019pqj,Gu:2020fem} (see also \cite{Kim:2021gyj,Sun:2021lsq} for additional related works). This provides an alternative route to the computation of the observables of interest, and in particular leads to a different expression for the elliptic genera of the BPS strings, which we find to be in perfect agreement with our 2d SQFT computation. On the other hand, we take advantage of the duality between the M-string SCFTs on a circle and 5d $\mathcal{N}=1^*$ $U(r)$ SYM. We specialize to the case $r=n=2$ and compute the partition function of the 5d gauge theory on the space $S^1\times\mathbb{C}^2/\mathbb{Z}_2$, building on and slightly adapting known results in 4d \cite{Gasparim:2009sns,Bonelli:2011jx,Bonelli:2011kv,Bruzzo:2013daa,Bruzzo:2014jza} which we uplift to five dimensions. Again we are able to match the resulting expressions with the partition function of the rank-2 M-string SCFT on $T^2\times\mathbb{C}^2/\mathbb{Z}_2$.

The ALE spaces $\mathbb C^2 / \mathbb Z_n$ are cones over the lens spaces $S^3/\mathbb Z_n$ and the latter have torsion at infinity. From this it follows that the Hilbert space of the system splits in selection sectors labeled by choices of a monodromy at infinity \cite{KronheimerNakajima,Douglas:1996sw}. This is reflected in the fact that there are many possible different M-string ALE partition functions, each labeled by one such choice. From the perspective of the BPS strings whose elliptic genera enter the ALE partition function, such a dependence is captured by the fact that the 2d CFTs governing the worldsheet theories of the strings become themselves relative theories. The field theoretical origin of this further discrete datum becomes manifest upon reduction of the 6d M-string SCFT on a circle: it is simply a choice of flat connection at infinty for the $U(r)$ gauge group that arises from the tensor multiplets. This choice of connection is is encoded in the choice of a morphism
\be
\boldsymbol{\omega}^{KK}_{UV}: \pi_1(S^3/\mathbb Z_n) \simeq \mathbb Z_n \to U(r).
\ee
Since we are in a Coulomb phase, the 5d gauge group is broken to its Cartan $U(1)^r$ and the choice of $\omega_{UV}$ translates to a choice of embedding 
\be
\boldsymbol{\omega}^{KK}_{IR}:\mathbb Z_n \to U(1)^r.
\ee
We show that the possible choices of boundary conditions at infinity, which are needed to fully specify the value of the M-string ALE partition function, are in 1-to-1 correspondence with the possible elliptic genera of the chiral 2d theories governing the dynamical properties of the solitons of the theory.

This work paves the way towards determining the ALE partition functions for more general 6d theories, by adopting the same strategy outlined above in terms of a string duality. In particular, a natural class of SCFTs whose ALE partition functions that are under current study are the E-string theories which arise by probing an M9 wall with a stack of M5 branes. Another important class of theories amenable to our approach are the so-called ``orbifolds of M-string'' theories \cite{Haghighat:2013tka}. The main complication which arises in 6d (1,0) SCFTs is that the tensorial abelian phases are in general supplemented by gauge degrees of freedom which form tensor-vector systems via a Green-Schwarz-Sagnotti mechanism \cite{Green:1984sg,Green:1984bx,Sagnotti:1992qw}. Because of this fact the discrete data which we have associated to the tensors in the form of a map $\text{Hom}(\mathbb Z_N, U(1)^r)$ is further dressed by choices of flat connection at infinity for the gauge groups in the tensor branch of the 6d (1,0) theories, giving rise to an interesting interplay between these further discrete choices and the structure of the solitonic BPS strings in those models. We plan to study this problem in future work \cite{wip}.

This paper is organized as follows:
In section \ref{sec:Mstring} we review basic features of the M-string SCFTs. In section \ref{sec:6dd} we place the SCFTs on a $T^2\times\mathbb{C}^2/\mathbb{Z}_n$ or $T^2\times \mathbf{TN}_n$ background: we begin in section \ref{sec:bg} by reviewing basic aspects of Taub-NUT \cite{Taub:1950ez,Newman:1963yy} and ALE geometries, and discuss the data that required to specify a background for M-string SCFTs on $T^2\times\mathbb{C}^2/\mathbb{Z}_n$; in section \ref{sec:IIBd} we discuss a Type IIB brane setup which is T-dual to the brane configuration that gives rise to the M-string SCFTs on on $T^2\times \mathbf{TN}_n$; finally, in section \ref{sec:bpsdof} we discuss the BPS degrees of freedom of the M-string SCFTs on $T^2\times\mathbb{C}^2/\mathbb{Z}_n$, which consist of the 2d $\mathfrak{su}(n)$ current algebra discussed above, of BPS particles localized on the individual M5/NS5 branes, and of BPS strings whose worldsheet theories $\QQ$ we determine. Section \ref{sec:Mtogether} is dedicated to computing the equivariant partition function of the M-string SCFTs on $T^2\times\mathbb{C}^2/\mathbb{Z}_n$. We determine the contributions arising from BPS particles in section \ref{sec:bpspc}, and the contributions from M-strings and chiral fermions in section \ref{sec:bpssc}; in section \ref{sec:comm} we then discuss some salient features of the partition function: its dependence on the topology of the spacetime geometry, its $\mathcal{N}=(2,0)$ limit where one recovers the Vafa-Witten partition function, and its relation to the partition function on the blown up space $T^2\times\widetilde{\mathbb{C}^2\times\mathbb{Z}_n}$ for which we conjecture a 6d version of the Nekrasov master formula which leads to an alternative expression for the M-string elliptic genus. Finally, in section \ref{sec:mstr} we analyze the properties of the 2d M-string SQFTs $\QQ$ in greater detail: in section \ref{sec:cc} we determine their `t Hooft anomalies and central charges, while in section \ref{sec:examples} we determine their elliptic genus explicitly in several examples and compare with the predictions from the Nekrasov master formula. In appendix \ref{sec:compar} we compare the rank 2 M-string partition function on $T^2\times\mathbb{C}^2/\mathbb{Z}_2$ to the partition function of the dual 5d $\mathcal{N}=1^*$ $U(2)$ gauge theory on the same spacetime, and find agreement between the two upon making some minor adjustments to the 5d gauge theoretic partition function, providing evidence for the duality and for the validity of our 6d partition function.

\section{A quick review of the M-string SCFTs}
\label{sec:Mstring}

\begin{figure}
\begin{center}
\resizebox{0.7\textwidth}{!}{
\begin{tikzpicture}
\begin{scope}[xshift=1.3 in, yshift=-1.2in]
  
 \shade[top color=blue!40, bottom color=red!10,xshift=-0.6in, rotate=90]  (0,-0.7) -- (2,-0.7) -- (2.6,-0.2) -- (0.6,-0.2)-- (0,-0.7);
   
 \draw[thick,xshift=-0.6in,rotate=90] (0,-0.7) -- (2,-0.7);
\draw[thick,xshift=-0.6in,rotate=90] (0,-0.7) -- (0.6,-0.2);
\draw[thick,xshift=-0.6in,rotate=90]  (0.6,-0.2)--(2.6,-0.2);
\draw[thick,xshift=-0.6in,rotate=90]  (2.6,-0.2)-- (2,-0.7);

\end{scope}

\begin{scope}[xshift=1.5 in, yshift=-1.2in]
  
 \shade[top color=blue!40, bottom color=red!10,xshift=-0.6in, rotate=90]  (0,-0.7) -- (2,-0.7) -- (2.6,-0.2) -- (0.6,-0.2)-- (0,-0.7);
   
 \draw[thick,xshift=-0.6in,rotate=90] (0,-0.7) -- (2,-0.7);
\draw[thick,xshift=-0.6in,rotate=90] (0,-0.7) -- (0.6,-0.2);
\draw[thick,xshift=-0.6in,rotate=90]  (0.6,-0.2)--(2.6,-0.2);
\draw[thick,xshift=-0.6in,rotate=90]  (2.6,-0.2)-- (2,-0.7);

\end{scope}

\begin{scope}[xshift=1.7 in, yshift=-1.2in]
  
 \shade[top color=blue!40, bottom color=red!10,xshift=-0.6in, rotate=90]  (0,-0.7) -- (2,-0.7) -- (2.6,-0.2) -- (0.6,-0.2)-- (0,-0.7);
   
 \draw[thick,xshift=-0.6in,rotate=90] (0,-0.7) -- (2,-0.7);
\draw[thick,xshift=-0.6in,rotate=90] (0,-0.7) -- (0.6,-0.2);
\draw[thick,xshift=-0.6in,rotate=90]  (0.6,-0.2)--(2.6,-0.2);
\draw[thick,xshift=-0.6in,rotate=90]  (2.6,-0.2)-- (2,-0.7);

\node at (-0.3,1.4){$\cdots$};

\node[below] at (-1, 0){$\qquad r$ M5};

\end{scope}

\begin{scope}[xshift=2.35 in, yshift=-1.2in]
  
 \shade[top color=blue!40, bottom color=red!10,xshift=-0.6in, rotate=90]  (0,-0.7) -- (2,-0.7) -- (2.6,-0.2) -- (0.6,-0.2)-- (0,-0.7);
   
 \draw[thick,xshift=-0.6in,rotate=90] (0,-0.7) -- (2,-0.7);
\draw[thick,xshift=-0.6in,rotate=90] (0,-0.7) -- (0.6,-0.2);
\draw[thick,xshift=-0.6in,rotate=90]  (0.6,-0.2)--(2.6,-0.2);
\draw[thick,xshift=-0.6in,rotate=90]  (2.6,-0.2)-- (2,-0.7);

\end{scope}

\draw[dashed] (0,-1.9) -- (6.9,-1.9);
\node at (6.9,-2.1){$\mathbf{TN}$};

\end{tikzpicture}
}
\end{center}

\caption{The six-dimensional (1,0) M-string theory of rank $r$ is the worldvolume theory of $r$ M5 branes probing a Taub-NUT space (upon decoupling the corresponding little string).}\label{fig:Mstrings}
\end{figure}
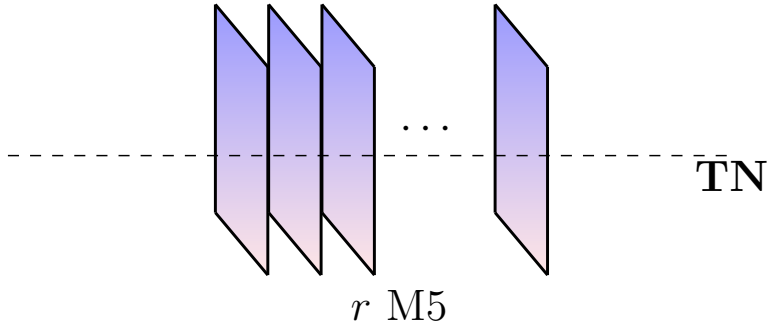

In this section, we will give a brief review of the six-dimensional superconformal field theories of interest in this paper: the M-string SCFTs \cite{Haghighat:2013gba}. These arise as the worldvolume theory of $r$ M5 branes, extended along directions $x_0,\dots,x_5$, probing a single centered Taub-NUT space $\mathbf{TN}$ extended along the directions $x_7,\dots,x_{10}$ --- see Figure \ref{fig:Mstrings}. The theories so obtained are little string theories; in this paper we are interested in the SCFT which is obtained via decoupling the little string mode from this system: we refer to the latter as the M-string SCFT of rank $r$.

We take the first two directions of the six-dimensional worldvolume, $x_0$ and $x_1$, to parametrize a torus $T^2$ of complex modulus $\tau$. For the moment we take $x_2,\dots,x_5$ to parametrize $\mathbb{C}^2$, but in later sections we will generalize this to the orbifold singularities of type $\mathbb{C}^2/\mathbb{Z}_{n}$ with either an ALE or a Taub-NUT metric.

It is well known that the rank $r$ M-string theory has a tensor branch that arises by spacing the $r$ M5 branes along direction $x_6$. On the tensor branch, the theory admits an effective low energy description in terms of $r$ tensor multiplets $T_a,$ for $a=1,\dots, r$ arising from the M5 branes. The bosonic components of the tensor multiplets are a two-form field $B^{(a)}$ and a real scalar $\phi^{(a)}$, whose vev $a^{(a)}$ parametrizes the position of the $a$-th M5 brane along $x_6$.

On top of this one needs to account for the degrees of freedom arising from the presence of the $\mathbf{TN}$ singularity. Naively, along the interval separating the $a$-th and $(a+1)$-th M5 brane the singularity gives rise to a 6d gauge algebra $\mathfrak{u}(1)^{(a)}$, while the two semi-infinite intervals give rise in 6d to a global symmetry algebra $\mathfrak{u}(1)^{(0)}\times \mathfrak{u}(1)^{(r)}$. However, the anti-diagonal combinations of $\mathfrak{u}(1)^{(a)} $ and  $\mathfrak{u}(1)^{(a+1)}$ for $a=0,\dots,r-1$ receive a mass by the Stuckelberg mechanism \cite{Berkooz:1996iz,Douglas:1996sw,Hanany:1997gh}, leaving behind an $\mathfrak{su}(1)^{r-1}$ gauge algebra and an unbroken $\mathfrak{u}(1)^{diag}$ global symmetry. The 6d SCFT possesses two dimensional BPS objects, the `M-strings', which arise from bound states of M2 branes stretched between parallel M5 branes, and can be thought of as point-like instantons for the 6d gauge algebra. A bound state of M-strings can be labeled by a vector of non-negative integers $\boldsymbol{\kappa}= (\kappa^{(1)},\dots,\kappa^{(r-1)})$, where $\kappa^{(a)}$ denotes the string's instanton charge with respect to $\mathfrak{su}(1)^{(a)}$.  The brane setup is summarized in the following table:
\be
\label{tab:M}
\begin{tabular}{r|ccccccc|cccc}
  & 0 & 1 & 2 & 3 & 4 & 5 & 6& 7 & 8 & 9 &10 \\
\hline
$r$ M5 & $\times$ & $\times$ & $\times$ & $\times$ & $\times$ & $\times$ &  &  &  & & \\
TN & $\times$ & $\times$ & $\times$ & $\times$ & $\times$ & $\times$ & $\times$ &  &  & & \\   
\hline
$\boldsymbol{\kappa}$ M2 & $\times$ & $\times$ &  &  &  & & & $\times$  &  & & \\
\end{tabular}
\ee

It is worth mentioning two alternative descriptions of this class of SCFTs that arise in different duality frames, which will be useful in what follows.\\

\paragraph{5d KK theory.} Let us compactify the M-theory setup along direction $x_1$. This leads to a stack of $r$ D4-branes probing the $\mathbf{TN}$ singularity in Type IIA. Their worldvolume is described by a mass deformation of the 5d $\mathcal N=2$ SYM theory, where we view the $\mathcal N=2$ vectormultiplet in the adjoint representation of $U(n)$ as a sum of an $\mathcal N=1$ adjoint vectormultiplet and a massive $\mathcal N=1$ adjoint hyper. We denote the resulting theory a 5d $\mathcal N=1^*$ theory, by analogy with the 4d $\mathcal N=2^*$ example. The positions of the D4 branes along direction $x^6$ parametrize the Coulomb branch of the theory, and the M2 branes descend to fundamental strings stretched between neighboring D4 branes. This setup is summarized in the following table.
\be
\label{tab:KK}
\begin{tabular}{r|cccccc|cccc}
  & 0 & 2 & 3 & 4 & 5 & 6& 7 & 8 & 9 &10 \\
\hline
$r$ D4 & $\times$ & $\times$ & $\times$ & $\times$ & $\times$ &  &  &  & & \\
TN & $\times$ & $\times$ & $\times$ & $\times$ & $\times$ & $\times$ &  &  & & \\   
\hline
$\boldsymbol{\kappa}$ F1 & $\times$&  &  &  & & & $\times$  &  & & \\
\end{tabular}
\ee

\paragraph{Type IIA brane setup.} We can also consider a different compactification of the M-theory setup. Namely, we compactify along the circle fiber of $\mathbf{TN}$. This gives rise to a set of $r$ NS5 branes (whose positions along direction $x_6$ parametrize the tensor branch), as well as a single D6 brane. The BPS strings in this duality frame are realized as bound states of D2-branes stretching between neighboring NS5-branes.  Schematically we have:
\be
\label{tab:IIA}
\begin{tabular}{r|cccccc|cccc}
  & 0 & 1 & 2 & 3 & 4 & 5 & 6& 7 & 8 & 9 \\
\hline
$r$ NS5 & $\times$ & $\times$ & $\times$ & $\times$ & $\times$ & $\times$ &  &  &  & \\
D6 & $\times$ & $\times$ & $\times$ & $\times$ & $\times$ & $\times$ & $\times$ &  &  & \\   
\hline
$\boldsymbol{\kappa}$ D2 & $\times$ & $\times$ &  &  &  & & $\times$  &  &  &\\
\end{tabular}
\ee

\noindent 

\section{M-strings on ALE/Taub-NUT backgrounds}

\label{sec:6dd}

In this section we take the spacetime of the M-string SCFT to be the space $T^2\times \mathbb{C}^2/\mathbb{Z}_n$ or its resolution $T^2\times\widetilde{\mathbb{C}^2/\mathbb{Z}_n}$, with either the $n$-center Taub-NUT or ALE metric. In section \ref{sec:bg} we provide some details on the background geometry. We then proceed in section \ref{sec:IIBd} to find a convenient duality frame to characterize the degrees of freedom contributing to the partition function.

\subsection{The background geometry}
\label{sec:bg}
The Taub-NUT space with $n$ centers can be viewed as a circle fibration
\begin{center}
\begin{tikzcd}
S^1 \arrow[r] & \mathbf{TN}_n \arrow[d, "\pi"] \\
& \mathbb{R}^3
\end{tikzcd}
\end{center}
and has metric
\be
ds^2 = R^2\left[V(\vec{x}) d\vec{x}^2+\frac{1}{V(\vec{x})}(d\theta + \beta)^2\right],
\ee
where $\vec{x}$ is the coordinate on $\mathbb{R}^3$ and $\theta$ is the angular coordinate along the circle fiber, which in this choice of coordinates has periodicity $4\pi$. The metric is  based on a Gibbons-Hawking ansatz with potential
\be
V(\vec{x}) = 1+\sum_{j=1}^n\frac{1}{\vert \vec{x}-\vec{x}_j\vert};
\ee
the one-form $\beta$ is determined by the condition
\be
d \beta = \star_{\mathbb{R}^3} d V.
\ee
The coordinates $\vec x_j$ determine the positions of the {centers} of $\mathbf{TN}_n$, where the circle shrinks to zero size. In particular, the charge 0 case corresponds simply to the geometry $\mathbb{R}^3 \times S^1$. The parameter $R$ denotes the asymptotic value of the radius of the circle fiber as $\vert\vec{x}\vert\to\infty$, which we denote by $S^1_\infty$. 
The $ALE$ limit is recovered by taking $R\to \infty$, while simultaneously rescaling $\vec{x}\to R^{-2} \vec{x}$ and $\vec{x}_j\to R^{-2}\vec{x}_j$. Taking then all $\vec{x}_j = 0$ one recovers the orbifold space $\mathbb{C}^2/\mathbb{Z}_n$ with an ALE metric, which we refer to as the $\mathbf{ALE}_n$ space below.\newline

\begin{figure}
    \centering
    \vspace{-.25in}
    \includegraphics[scale=0.45]{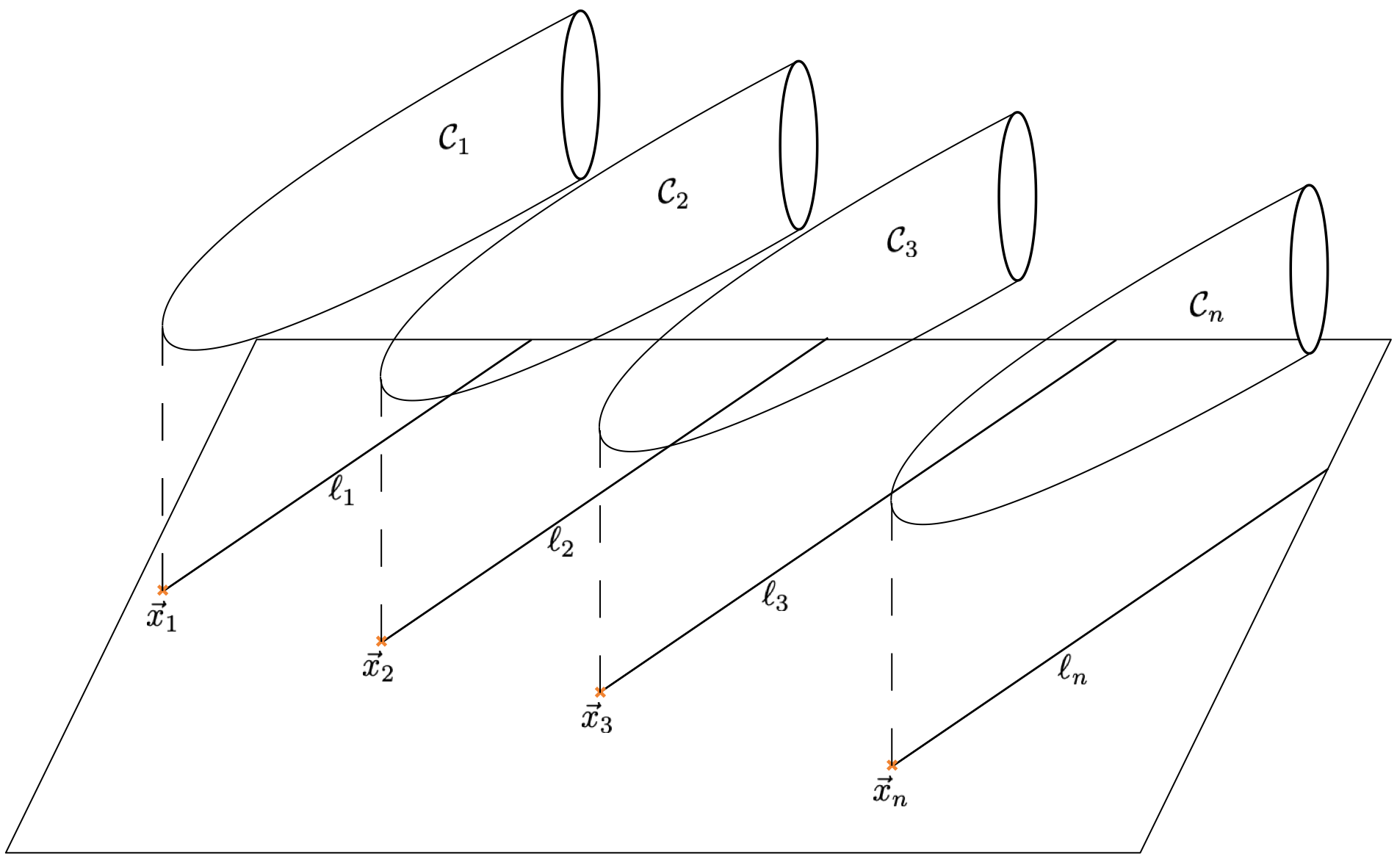}
    \caption{Basis of non-compact two-cycles $\mathcal{C}_i$ on $\mathbf{TN}_n$.}
    \label{fig:elli}
\end{figure}

The second homology group of $\mathbf{TN}_n$ and $\mathbf{ALE}_n$ is isomorphic to the weight lattice of $A_{n-1}$. Before discussing its generators, it is convenient to introduce the following set of circle bundles \cite{Witten:2009xu}:
\be
\mathcal{C}_i = \pi^{-1}(\ell_i)
\ee
over semi-infinite intervals that originate at the $\vec{x}_i$ and extend to infinity, as in figure \ref{fig:elli}.
The difference $\ell_{i}-\ell_{i+1}$ is homologous to an interval starting at $\vec{x}_i$ and ending at $\vec{x}_{i+1}$. Let us now define the following set of two-cycles:
\be
\Sigma_i = \pi^{-1}(\ell_i-\ell_{i+1})\qquad i=1,\dots,n-1
\ee
which are topologically two-spheres. This collection of compact two-cycles form a basis of generators for $H_2(\mathbf{TN}_n)$; they intersect according to the negative of the Cartan matrix of $A_{n-1}$, which we denote by $C^{A_{n-1}}_{ij}$:
\be
\Sigma_i\cdot \Sigma_j = -C^{A_{n-1}}_{ij} =
\begin{cases}
-2 & \text{ if } i=j \\
1 &  \text{ if } \vert i-j\vert = 1\\
0 & \text{ otherwise}.
\end{cases}
\ee
It is also convenient to consider the following additional two cycle:
\be
\Sigma_0 = \mathcal{C}_n-\mathcal{C}_1
\ee
which has self-intersection $-2$, nonvanishing intersection $1$ with $\Sigma_1$ and $\Sigma_{n-1}$, all intersection 0 with $\Sigma_2,\dots,\Sigma_{n-2}$. In other words, the set of two cycles $\Sigma_0,\dots,\Sigma_{n-1}$ has self-intersection matrix given by minus the Cartan matrix $C^{\widehat{A}_{n-1}}$ of affine $A_{n-1}$. In the $\mathbf{ALE}_n$ limit where $R\to\infty$, the two-cycles $\Sigma_{1},\dots,\Sigma_{n-1}$ can be identified with the exceptional divisors that arise from blowing up the orbifold singularity $\mathbb{C}^2/\mathbb{Z}_n$.

In what follows, it will also be convenient to exploit the fact that there exists a natural collection of $n$ line bundles $\mathcal{L}_1,\dots,\mathcal{L}_n$ on $\mathbf{TN}_n$ (see \cite{Witten:2009xu,Dijkgraaf:2007sw} for an explicit construction) with anti-selfdual curvature $\alpha_i$ such that
\be
\int_{\mathcal{C}_j} \frac{\alpha_i}{2\pi} = \delta_{ij}.
\ee
The tensor product $\mathcal{L}_* = \otimes_{i=1}^n \mathcal{L}_i$ can be assigned a globally well-defined connection $\Lambda$, so for an arbitrary real parameter $t$ it is possible to define a line bundle $\mathcal{L}_*^t$ with connection $\Lambda_t = t\Lambda$.

A generic unitary line bundle on $\mathbf{TN}_n$ with anti-selfdual connection may then be expressed as a tensor product of the line bundles mentioned above:
\be
\mathcal{L} =  \mathcal{L}_*^t\otimes(\bigotimes_{i=1}^n \mathcal{L}_i^{p_i}),
\label{eq:line}
\ee
which is determined by a tuple of integers $p_i$ and a real number $t$, modulo the equivalence 
\be
(t,p_1,\dots,p_n)\sim (t+1,p_1-1,\dots,p_n-1).
\ee
The connection associated to a line bundle can possess nontrivial holonomy along the circle at infinity of $\mathbf{TN}_n$. This is not the case for the line bundles $\mathcal{L}_i$, which have trivial holonomy as their connection vanishes at infinity. On the other hand, one can show that $\mathcal{L}_*^t$ with the globally well-defined connection $\Lambda_t$ has holonomy $\exp(2\pi i t)$.

Finally, in Type IIA string theory one may turn on a nontrivial NS two-form background; this gives rise to parameters
\be
\theta_i = \int_{\mathcal{C}_i} B^{NS},\qquad i=1,\dots,n,
\label{eq:th}
\ee
which are periodic modulo $2\pi$ and are defined up to overall shifts $(\theta_1,\dots,\theta_n) \to (\theta_1+\int_{S^1_\infty} \lambda,\dots,\theta_n+\int_{S^1_\infty}\lambda)$ which arise from gauge transformations of the two-form field $B^{NS} \to B^{NS} + d \lambda$. We choose to turn on a background such that no two $\theta_i$ are identical. Without loss of generality, we may assume that $\theta_1<\theta_2<\dots<\theta_n$ (mod $2\pi$), since our choice of $B^{NS}$ guarantees that we may rearrange the positions $\vec{x}_i$ without encountering a singularity in the complexified K\"ahler moduli space.
\newline

In the ALE limit, the two-form associated to the line bundle $\mathcal{L}_*$ becomes non-normalizable, and a generic unitary anti-selfdual line bundle can be expressed as a tensor product
\be
\mathcal{L} = \bigotimes_{i=1}^n \mathcal{L}_i^{p_i},
\ee
where the $p_i$ are integers modulo the equivalence $(p_1,\dots,p_n)\sim (p_1-1,\dots,p_n-1)$. Since the asymptotic boundary of $\mathbf{ALE}_n$,
\be
\partial \mathbf{ALE}_n = S^3/\mathbb{Z}_n
\ee
has first homotopy group $\mathbb{Z}_n$, line bundles can have non-trivial monodromy of infinity. From the explicit form of the connection in the ALE limit $R\to\infty$, one can verify that all line bundles $\mathcal{L}_i$  have monodromy $\exp(-2\pi i /n)$. Given the $\mathcal{L}_j$, one may construct the following line bundles:
\be
\mathcal{R}_j = \bigotimes_{p=1}^j\mathcal{L}_p^{-1},\qquad j=1,\dots,n-1,
\ee
whose monodromy is given by $e^{2\pi i j/n}$.
On $\mathbf{ALE}_n$, the two-forms $c_1(\mathcal{R}_i)$ can be shown to form a basis of $H_2(ALE_n,\mathbb{Z})$ which is dual to the basis $\Sigma_1,\dots\Sigma_{n-1}$ of section \ref{sec:bg} \cite{Douglas:1996sw}. One may therefore expand the first Chern class of an arbitrary bundle $\mathcal{L}$ on $ALE_n$ as
\be
\label{eq:c1}
c_1(\mathcal{L}) = \sum_{j=1}^{n-1}u_j c_1(\mathcal{R}_j)
\ee
for a set of integers $(u_1,\dots,u_{n-1})\in\mathbb{Z}^{n-1}$.\newline

In this paper we are interested in performing computations in an equivariant setup. Both the $\mathbf{TN}_n$ and $\mathbf{ALE}_n$ spaces have a triholomorphic $U(1)$ isometry (which acts by shifts on $\theta$); they also possess an additional $U(1)$ isometry when the centers are aligned, which we assume to be the case. The non-triholomorphic $U(1)$ isometry enhances to $SU(2)$ in the singular limit where all centers coincide. In the orbifold limit $\mathbb{C}^2/\mathbb{Z}_n$, we can identify the chemical potentials as follows: denote by $z_1,z_2$ the complex coordinates on $\mathbb{C}^2$. The $SU(2)_t\times SU(2)_x$ isometry acts on these coordinates as:
\be
(z_1,z_2) \to (t x z_1,t x^{-1} z_2),
\ee
where $t = e^{2\pi i \epsilon_+}$ and $ x = e^{2\pi i \epsilon_-}$. We embed $\mathbb{Z}_n$ into $SU(2)_x$, so that its action on coordinates is given by
\be
(z_1,z_2) \to (e^{2\pi i/n} z_1,t e^{-2\pi i/n} z_2).
\ee
In the orbifolded space, the chemical potential for the triholomorphic $U(1)$ isometry is identified with $\epsilon_-$, while the one for the non-triholomorphic $SU(2)$ isometry is identified with $\epsilon_+$. For convenience we also report here the value of the equivariant parameters $\epsilon_+[j],\epsilon_-[j]$ for $j=1,\dots,n$ at the $n$ fixed points of the $T^2$ action on the resolved $\mathbf{ALE}_n$ space \cite{Bonelli:2012ny}:
\bea
\epsilon_+[j]=\epsilon_+,\qquad \epsilon_-[j] = n\,\epsilon_-+(n+1-2j)\epsilon_+.
\label{eq:eqv}
\eea
\newline
Having reviewed basic aspects of the geometries of interest, let us now discuss the gauge bundle data needed to specify a background for the M-string SCFTs on the tensor branch.\footnote{ Our exposition here is schematic; we refer to \cite{wip} for a more detailed treatment.} This includes prescribing a monodromy at infinity for the gauge fields, including the background gauge fields. On the one hand, the $\mathfrak{u}(1)^{diag}$ background gauge field, which arises from the D6 brane, can be trivialized by a B-field gauge transformation \cite{Witten:2009xu} while simultaneously performing a cyclic relabeling of the $\mathcal{C}_i$, so its monodromy can be set to zero without loss of generality. On the other hand, different choices of monodromy for the two-form fields $B^{(a)}$ in the tensor multiplets do determine physically inequivalent backgrounds. A convenient way to characterize these choices is in the 5d KK picture, where a two-form field $B^{(a)}$ translates to a one-form connection $A^{KK,(a)}$ for an $U(1)$ gauge bundle $\mathcal{V}^{(a)}$, whose monodromy at infinity is characterized by an irreducible representation of 
\be
\text{Hom}(\pi_1(S^3/\mathbb{Z}_n),U(1))\simeq \mathbb{Z}_n,
\ee
or equivalently by a phase 
\be
e^{2\pi i\frac{\omega^{KK,(a)}}{n}}, \qquad \omega^{KK,(a)}\in\mathbb{Z}_n.
\ee
Once the background is specificed by the choice of monodromy, inequivalent field configurations can be considered which are distinguished by the instanton number
\be
n^{KK,(a)} = \int_{\mathbf{ALE}_n} ch_2(\mathcal{V}^{(a)})
\ee
and by the amount of flux for each of the gauge groups $U(1)^{KK,(a)}$ through the two-cycles of $\mathbf{ALE}_n$. We may expand the gauge bundle $\mathcal{V}^{(a)}$ in terms of the bundles $\mathcal{R}_j$ as follows:
\be
\mathcal{V}^{(a)} = \otimes_{j=1}^{n-1} \mathcal{R}_j^{u^{KK,(a)}_j}.
\ee
Requiring $\mathcal{V}^{(a)}$ to possess the correct monodromy imposes the following constraint:
\be
\sum_{j=1}^{n-1}j u_j^{KK,(a)} = \omega^{KK,(a)}_j \qquad \text{mod } n.
\ee
Then, the first Chern class
\be
c_1(\mathcal{V}^{(a)}) = \sum_{j=1}^{n-1} u^{KK,(a)}_j c_1(\mathcal{R}_j), 
\ee
encodes the fluxes for gauge group $U(1)^{KK,(a)}$,
\be
u^{KK,(a)}_j = \int_{\Sigma_j}c_1(\mathcal{V}^{(a)}) \in \mathbb{Z},
\ee
through the compact two-cycles of $\mathbf{ALE}_n$.

\subsection{Type IIB description}
\label{sec:IIBd}

\begin{figure}
    \centering
    \includegraphics[width=\textwidth]{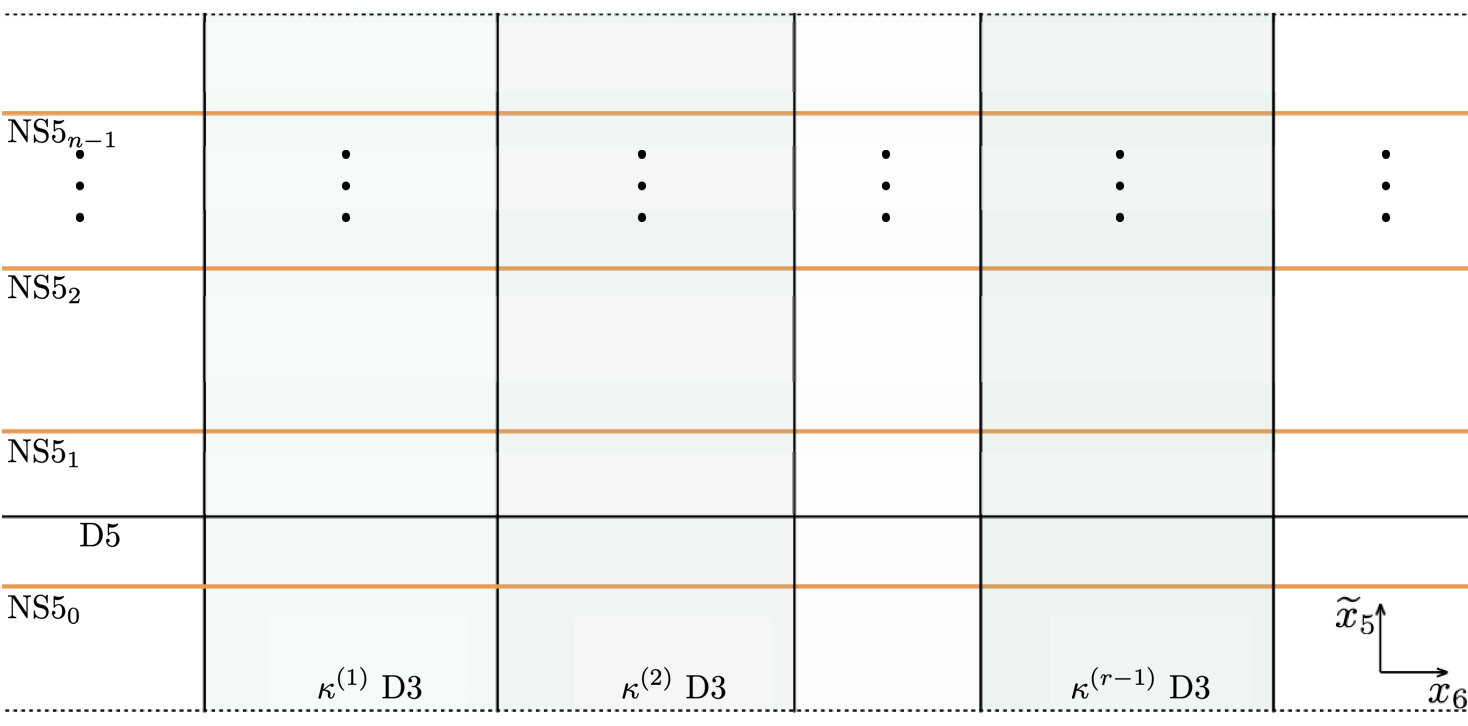}
    \caption{Type IIB brane setup corresponding to the M-string SCFT on $T^2\times\mathbf{TN}_n$. The figure displays a BPS brane configuration corresponding to a bound state of $\boldsymbol{\kappa}$ M-strings. The vertical direction is identified along the dashed lines and corresponds to the dual to the Taub-NUT circle.}
    \label{fig:20config}
\end{figure}

Consider now the rank $r$ M-string theory on $T^2\times \mathbf{TN}_n$, from which we will eventually take the limit $T^2\times \mathbf{ALE}_n$. In this section we aim to understand better the degrees of freedom contributing the partition function on the latter background. We do this by exploiting a dual realization in Type IIB string theory, which we arrive at by performing T-duality  along the circle of $\mathbf{TN}_n$, starting from the following Type IIA setup:
\be
\label{tab:IIATN}
\begin{tabular}{r|cccccc|cccc}
  & 0 & 1 & 2 & 3 & 4 & 5 & 6& 7 & 8 & 9 \\
\hline
$\mathbf{TN}_n$ & $\times$ & $\times$ &&&&&  $\times$ & $\times$ & $\times$ & $\times$\\
$r$ NS5 & $\times$ & $\times$ & $\times$ & $\times$ & $\times$ & $\times$ &  &  &  & \\
D6 & $\times$ & $\times$ & $\times$ & $\times$ & $\times$ & $\times$ & $\times$ &  &  & \\   
\hline
$\boldsymbol{\kappa}$ D2 & $\times$ & $\times$ & & & & & $\times$ & & & \\
\end{tabular}
\ee
Under T-duality, the Taub-NUT space is replaced by a collection of $n$ NS5 branes on $\mathbb{R}^3\times S^1$, where $S^1$ is the dual to the Taub-NUT circle and has radius $1/R$. The positions of the NS5 branes along $\mathbb{R}^3$ coincide with the positions $\vec{x}_i$ of the Taub-NUT centers on the Type IIA side; on the other hand, the angular positions on $S^1$ are given by the parameters $\theta_i$ defined in equation \eqref{eq:th}, and the gauge transformation shifting all $\theta_i\to \theta_i+\int_{S^1_\infty}\lambda$ corresponds to the freedom to rotate the dual $S^1$ by the same angle $\int_{S^1_\infty}\lambda$. In what follows, we keep the $\theta_i$ to be distinct but set all $\vec{x}_i\to 0$.\newline After T-duality, we end up with the following brane setup:
\be
\label{tab:20IIB}
\begin{tabular}{r|ccccc|cc|ccc}
  & 0 & 1 & 2 & 3 & 4 & $\tilde{5}$ & 6& 7 & 8 & 9 \\
\hline
$n$ NS5 & $\times$ & $\times$ & & & &  & $\times$ &$\times$ & $\times$ & $\times$ \\ 
$r$ NS5 & $\times$ & $\times$ & $\times$ & $\times$ & $\times$ & $\times$ &  &  &  & \\
D5 & $\times$ & $\times$ & $\times$ &  $\times$& $\times$ &  & $\times$ &  &  & \\   
\hline
$\boldsymbol{\kappa}$ D3 & $\times$ & $\times$ &  &  & &$\times$  &$\times$ &   && \\
\end{tabular}
\ee
We take coordinates $x_2,x_3,$ and $x_4$ to parametrize $\mathbb{R}^3$ and  $\tilde{x}_5$ to parametrize the T-dual of the Taub-NUT circle. The configuration is illustrated in figure \ref{fig:20config}, where directions $\tilde{x}_5$ and $x_6$ are displayed and the $n$ horizontal NS5 branes are the T-dual counterpart of the $\mathbf{TN}_n$ singularity. The ALE limit corresponds to shrinking the circle parametrized by $\tilde{x}_5$ to zero size. The D6 brane becomes a D5 brane, and BPS strings are now represented by D3 branes which end on the $r$ vertical NS5 branes.

\subsection{BPS degrees of freedom}
\label{sec:bpsdof}
The Type IIB brane configuration discussed above gives rise to three types of BPS degrees of freedom which are dynamical in the 6d SCFT. We discuss them in turn:
\paragraph{BPS particles.} Each of the NS5 branes contributes a 6d $\mathcal{N}=(1,0)$ tensor multiplet and a hypermultiplet to the BPS spectrum. Upon compactification to 5d, these give rise respectively to the massless abelian vector multiplets and neutral hypermultiplets in $\mathcal{N}=1^*$ $U(r$) SYM on the Coulomb branch, as well as their KK towers.
\paragraph{Chiral fermions.} M5 branes on the spacetime $T^2\times\mathbf{TN}_n$ are known to possess BPS degrees of freedom that are localized at the Taub-NUT center and give rise to chiral fermionic particles on $T^2$ \cite{Dijkgraaf:2007sw}. These degrees of freedom can also be understood within the present Type IIB frame. Indeed, the two-dimensional intersection between the two stacks of NS5-branes is known \cite{Itzhaki:2005tu} to support a system of $n\cdot r$ complex fermions which transform in the bifundamental representation of $U(n)\times U(r)$. The presence of the D5 brane can be ignored in the present context. Let us assume for the moment that all NS5 branes in each of the two stacks are coincident. In other words, let us look at the rank $r$ SCFT at the superconformal fixed point, with no $B$-field turned on at the singularity.  As detailed in \cite{Dijkgraaf:2007sw}, the fermions furnish a $\mathfrak{u}(n\,r)_1$ current algebra, and the symmetries supported on the two stacks of NS5 branes are visible in the conformal embedding
\begin{equation}
\label{eq:confemb}
    \mathfrak{u}(n\,r)_1
    \supset
    \mathfrak{su}(r)_n\times \mathfrak{su}(n)_r\times \mathfrak{u}(1)_{n\,r}.
\end{equation}
The abelian factor $\mathfrak{u}(1)_{n\, r}$ corresponds to the anti-diagonal combination of the abelian factors of $U(n)$ and $U(r)$, while the diagonal combination is decoupled from the fermions. From the system of free fermions, one can reproduce \cite{Dijkgraaf:2007sw} the celebrated result of Vafa and Witten \cite{Vafa:1994tf} that the twisted partition function of $\mathcal{N}=4 $ $ U(r)$ SYM on $\mathbb{C}^2/\mathbb{Z}_n$ is captured in terms of an affine $\mathfrak{su}(n)$ algebra at level $r$: in this setup, the $SU(r)$ symmetry is treated as dynamical, so one is led to consider the coset
\begin{equation}
\mathfrak{u}(n\, r)_1/\mathfrak{su}(r)_n
\end{equation}
which indeed contains a factor of $\mathfrak{su}(n)_r$. The extra $\mathfrak{u}(1)_{n\,r}$ factor that appears in \eqref{eq:confemb} can be interpreted as a compact chiral boson on a circle of radius $\sqrt{2 n r}$. As pointed out in \cite{Dijkgraaf:2007sw}, since the Taub-NUT circle has asymptotic radius at infinity there exist field configurations corresponding to monopoles going around the Taub-NUT circle which do not arise in the ALE limit where the asymptotic circle is decompactified. In this limit it is known \cite{Belavin:2011pp, Nishioka:2011jk, Belavin:2011tb} that the relevant algebra gets replaced by $\mathcal{H}\times \mathfrak{su}(n)_r$, where $\mathcal{H}$ is the Heisenberg algebra associated to a non-compact free boson.

Let us now deform the 6d SCFT by moving to the tensor branch. This corresponds to separating the $n\, r$ chiral fermions into $r$ sets of $n$. At the level of current algebra, it is reflected by the conformal embedding
\be
    \mathfrak{u}(n)_1^r
    \subset
    \mathfrak{u}(n\, r)_1.
\ee
For each M5 brane, passing from Taub-NUT to ALE corresponds to replacing $ \mathfrak{u}(n)_1 $ with $\mathcal{H}\times\mathfrak{su}(n)_1$. All in all, we are left with a $\mathfrak{su}(n)^r$ current algebra and $r$ copies of the Heisenberg algebra.

\paragraph{BPS strings.}
In the setup of figure \ref{fig:20config}, the degrees of freedom can be obtained by reducing the worldvolume theory of bound states of D3 branes down to 2d. This setup is, in fact, a special case of a more general class of configurations that have been studied in \cite{Hanany:2018hlz}. The D3 branes give rise to 2d $\mathcal{N}=(0,4)$ quiver gauge theories which have been dubbed (0,4) \emph{brane box models}, which in our specific setup we interpret as the 2d $\mathcal{N} = $ (0,4) quiver gauge theories $\QQTN$ describing bound states of M-strings probing the Taub-NUT space. 

\begin{figure}
    \centering
    \includegraphics[width=\textwidth]{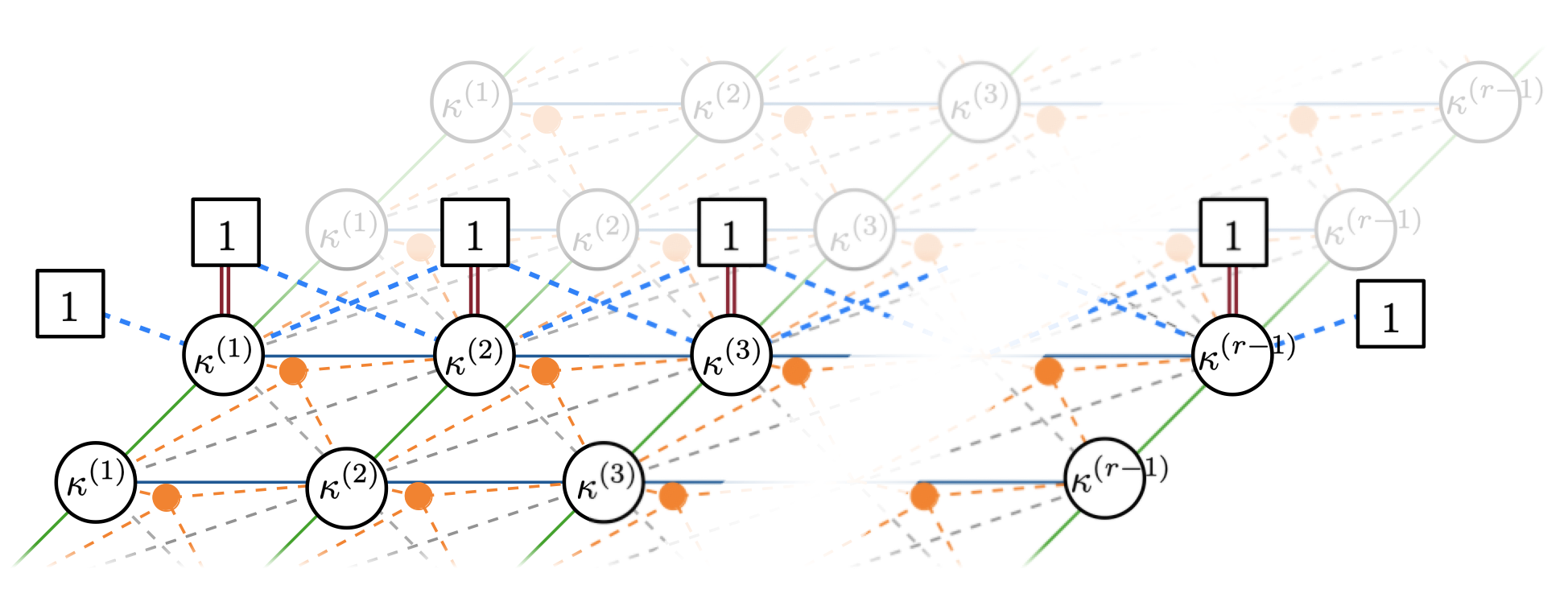}
    \caption{The 2d $\mathcal{N}=(0,4)$ quiver gauge theory $\QQTN$ which captures the degrees of freedom of a bound state of $\boldsymbol{\kappa}=(\kappa^{(1)},\dots,{\kappa}^{(r-1)})$ strings of the rank-$r$ M-string SCFT on $T^2\times\mathbf{TN}_n$. The degrees of freedom associated to the nodes and edges of the quiver are discussed in the main text and in figures \ref{fig:plaqj} and \ref{fig:plaq0}.}
    \label{fig:quivv}
\end{figure}

\begin{figure}
\begin{center}
\includegraphics[width=\textwidth]{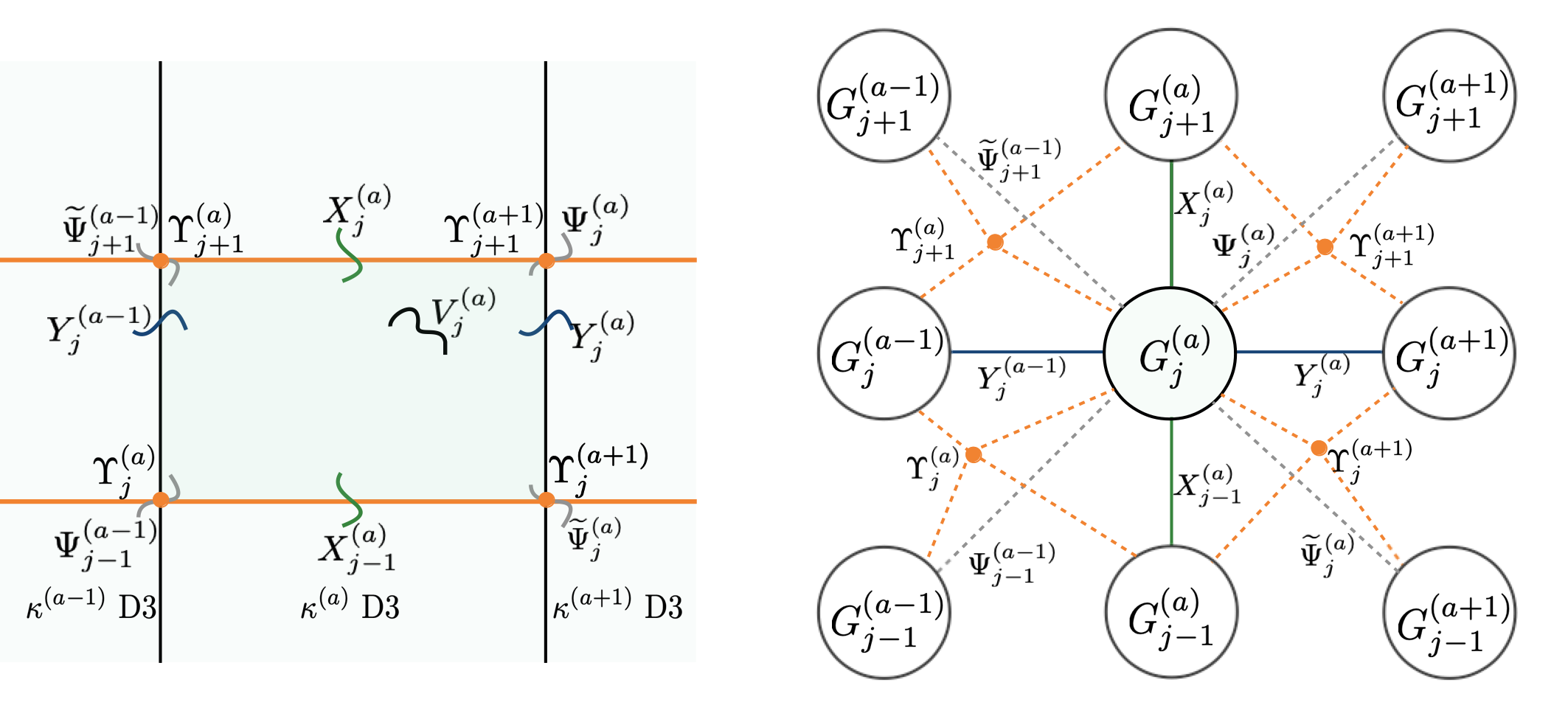}
\caption{ 
Fields that couple to the $(a,j)$ D3 brane (highlighted in light green) for $j\neq 0$, in the Type IIB brane setup. Here, $G^{(a)}_0=\dots=G^{(a)}_{j-1}=U(\kappa^{(a)})$. On the left side of the figure, the strings ending on the $(a,j)$ D3 brane are shown schematically. On the right side of the figure, the gauge node that corresponds to the $(a,j)$ D3 brane, as well as the fields that couple to it and the neighboring nodes are shown. Hyper- and Fermi multiplets are depicted respectively by continuous and dashed lines. The tetravalent Fermi multiplets are represented by the orange dots.
}
\label{fig:plaqj}
\end{center}
\end{figure}

\begin{figure}
\begin{center}
\includegraphics[width=\textwidth]{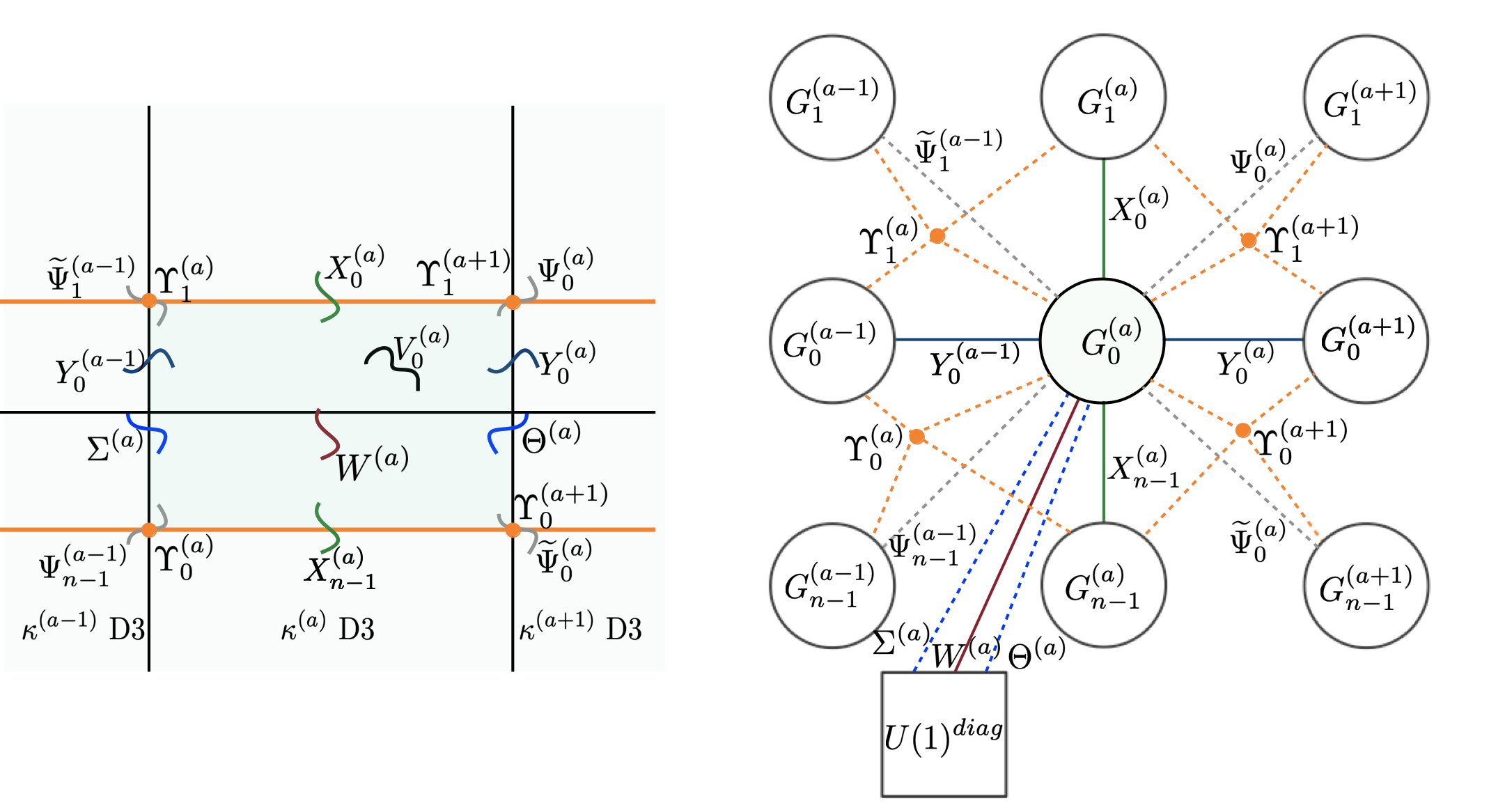}
\caption{Fields that couple to the $(a,0)$ D3 brane in the Type IIB brane setup. We take $G^{(a)}_0=\dots=G^{(a)}_{j-1}=U(\kappa^{(a)})$.}
\label{fig:plaq0}
\end{center}
\end{figure}

The quiver $\QQTN$ is depicted in figure \ref{fig:quivv} and consists of $(r-1)\cdot n$ nodes associated to the gauge groups $G^{(a)}_j = U(\kappa^{(a)})$, where $G^{(a)}_j$ arises from the stacks of $\kappa^{(a)}$ coincident D3 branes suspended between the $a$ and $a+1$-st vertical and the $j$ and $j+1$-st horizontal NS5 branes, as depicted on the left sides of figures \ref{fig:plaqj} and \ref{fig:plaq0}. The degrees of freedom supported on the D3 branes corresponding to the $(a,j)$ node give rise to a number of $\mathcal{N}=(0,4)$ multiplets:
\begin{itemize}
\item[-] Vector multiplets $V^{(a)}_j$ for gauge groups $G^{(a)}_j$;
\item[-] Twisted hypermultiplets $X^{(a)}_{j}$ in the bifundamental representation of $G^{(a)}_{j}\times G^{(a)}_{j+1}$;
\item[-] Hypermultiplets $Y^{(a)}_{j}$ in the bifundamental representation of $G^{(a)}_{j}\times G^{(a+1)}_{j}$;
\item[-] Fermi multiplets $\Psi^{(a)}_{j},\widetilde\Psi^{(a)}_{j},$ respectively in the bifundamental representation of $G^{(a)}_{j}\times G^{(a+1)}_{j + 1}$ and of $G^{(a)}_{j}\times G^{(a+1)}_{j - 1}$.\footnote{ We take the on-shell degrees of freedom of a $(0,4)$ Fermi multiplet to consist of one complex chiral fermion.}
\end{itemize}
These degrees of freedom are displayed in figure \ref{fig:plaqj}. Additionally, at the $(a,0)$ node of the quiver the D3-D5 strings give rise to the following 2d $(0,4)$ multiplets:
\begin{itemize}
\item[-] Hypermultiplets $W^{(a)}$ in the bifundamental representation of $G^{(a)}_0\times U(1)^{(a)}$;
\item[-] Fermi multiplets $\Sigma^{(a)}$ and $\Theta^{(a)}$, respectively in the bifundamental representation of $G^{(a)}_0\times U(1)^{(a-1)}$ and $U(1)^{(a+1)}\times G^{(a)}_0$;
\end{itemize}
recall that the $U(1)^{(a)}$ are all identified by the Stuckelberg mechanism. The degrees of freedom that couple to the $(a,0)$ nodes are displayed in figure \ref{fig:plaq0}.\\


The degrees of freedom mentioned thus far give rise to a quiver $\mathcal{Q}^{anom}_{\boldsymbol{\kappa}}$ that suffers from gauge anomalies and in itself is inconsistent \cite{Hanany:2018hlz}. Specifically, the anomaly affects the abelian factors of the $G^{(a)}_j=U(\kappa^{(a)})$ gauge groups. The Type IIB brane construction provides a clue as to the mechanism that ensures the vanishing of this gauge anomaly. Namely, the intersections between the $a$-th vertical and $j$-th horizontal NS5 branes support complex chiral fermion, which supplies an additional Fermi multiplet $\Upsilon^{(a)}_{j}$ to the 2d quiver gauge theory. These are precisely the BPS chiral fermions we encountered earlier in this section. The authors of \cite{Hanany:2018hlz}, building on \cite{Costello:2018fnz}, argue that the Fermi multiplet $\Upsilon^{(a)}_j$ transforms in the product of the determinant representations of $G^{(a)}_j$ and $ G^{(a-1)}_{j-1}$ and of the anti-determinant representations of  $G^{(a-1)}_{j-1}$ and $ G^{(a)}_{j-1}$. This is represented in figure \ref{fig:quivv} by a tetravalent node:
\begin{center}
\includegraphics[width=0.27\textwidth]{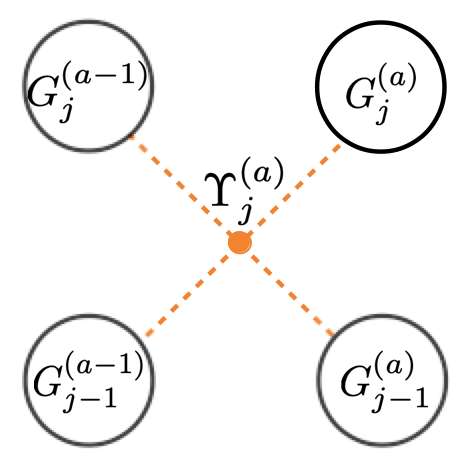}
\end{center}
Indeed, for a single chiral fermion localized at an intersection between the $a$ and $j$ NS5 branes we can write a coupling
\be
\int_{T^2} J^{(a)}_j(\Tr{A}^{(a)}_{j}-\Tr{A}^{(a-1)}_{j}+\Tr{A}^{(a-1)}_{j-1}-\Tr{A}^{(a)}_{j-1})
\label{eq:cpl}
\ee
between the fermion's $U(1)^{(a)}_j$ current and the abelian component of the gauge fields on the D3 branes that end on the $(a,j)$ intersection. Including this coupling has the effect of canceling the abelian gauge anomalies and making the resulting theory well-defined \cite{Hanany:2018hlz}.\\

As discussed earlier in this section, on the tensor branch the current algebra is broken from $\mathfrak{u}(n r)_1$ to $\mathfrak{u}(n)^r_1$. Recall that in order to arrive at the Type IIB picture we have replaced the space $\mathbb{C}^2/\mathbb{Z}_n$ by a Taub-NUT space of charge $n$, and as a consequence the Type IIB picture includes spurious degrees of freedom which do not arise in the genuine ALE background. Specifically, while in the Taub-NUT case we found a collection of $n$ complex chiral fermions localized on each of the $r$ vertical NS5 branes, giving rise to a 
\be
\mathfrak{u}(n)_1^r\supset (\mathfrak{u}(1)_n \oplus \mathfrak{su}(n)_1)^r
\ee
current algebra, in the ALE case the relevant chiral algebra is
\be
(\mathcal{H} \oplus \mathfrak{su}(n)_1)^r,
\label{eq:relchi}
\ee
which is obtained by replacing the abelian factors by $r$ copies of the Heisenberg algebra. It is useful to note that the linear combinations of currents $\sum_{j=0}^{n-1}J^{(a)}_j$ for $a=1,\dots, r$, which correspond to the abelian factors in $\mathfrak{u}(n)_1^r$, are decoupled from the gauge fields on the D3 branes. As a consequence, cancelation of gauge anomalies in the ALE case can be accomplished by taking $J^{(a)}_0,\dots, J^{(a)}_{n-1}$, subject to $\sum_{j=0}^{n-1}J^{(a)}_j = 0$, to be the currents in the Cartan of the $a$-th copy of $\mathfrak{su}(n)_1$ in equation \eqref{eq:relchi}, and coupling them to the gauge fields of the quiver $\mathcal{Q}^{anom}_{\kappa}$ according to equation \eqref{eq:cpl}.

The novel feature of the ALE chiral algebra \eqref{eq:relchi} is that each of the $r$ factors of $\mathfrak{su}(n)_1$ possesses $n$ distinct integrable highest weight representations, which are characterized by a choice of an $n$-dimensional affine  Dynkin label $(0,\dots,0,1,0,\dots 0)$, with a single nonzero entry associated to the $\omega^{(a)}$-th fundamental weight, where $\omega^{(a)} \in\{0,\dots,n-1\}$. The dependence on this discrete data is natural: the two-form field $B^{(a)}$ on the $a$-th NS5 brane reduces in 5d to a gauge field $U(1)^{KK,(a)}$ for which one must specify a choice of monodromy $e^{2\pi i \frac{\omega^{KK,(a)}}{n}}$ on the boundary of ALE. The authors of \cite{Dijkgraaf:2007sw} show that the choice of a monodromy $\omega^{KK,(a)}$ at infinity identifies a choice of integrable highest weight representation for $\mathfrak{su}(n)_1$ such that that
\be
\omega^{(a)} = \omega^{KK,(a)},
\ee
which is essentially a realization of the McKay correspondence.

To summarize, we found that the 2d (0,4) quiver gauge theory describing instanton strings in the $T^2\times \mathbf{ALE}_n$ background is obtained by coupling the degrees of freedom arising from the D3 branes to $r$ copies of the Heisenberg algebra as well as to a chiral WZW theory with algebra $\mathfrak{su}(n)_1$. We write the resulting system as
\be
\QQ = \mathcal{H}^r \times (\mathfrak{su}(n)_1 \ltimes \mathcal{Q}^{anom}_{\boldsymbol{\kappa}}).
\ee
The theory so defined is indeed relative and does not possess a partition function. Rather, it has a vector of conformal blocks, which on $T^2$ can be interpreted as the components of a vector valued meromorphic Jacobi modular form --- see the discussion in section \ref{sec:cc} for more details. 
The relative nature of the 2d theory was already observed by Itzhaki, Kutasov and Seiberg in the context of the I-brane configuration \cite{Itzhaki:2005tu} --- in the duality frame discussed in section \ref{sec:IIBd} we are decorating the original I-brane configuration with D3 branes. The fact that the 2d BPS strings acquire a richer global structure, encoded in a bulk-boundary system, allows for the overall sum over BPS sectors to have the necessary extra dependence on choices of discrete data at infinity. As already remarked by the authors of  \cite{Dijkgraaf:2007sw} albeit in a non-equivariant setting, we notice that the choices that specify a flat connection at infinity for the 5d $U(r)$ $\mathcal N=1^*$ KK theory in its Coulomb phase, namely a morphism $\text{Hom}(\mathbb Z_n,U(1)^r)$, are indeed in 1-to-1 correspondence with the choices of integrable irreducible highest weights needed to specify a number from the partition vector of the 2d relative CFT on the BPS string worldsheets.\\

One may wish to consider the possibility of generalizing the quiver by allowing the gauge groups $U(\kappa^{(a)}_j)$ for a fixed $a$ to have different ranks, which would correspond to more general BPS string sectors. However, cancelation of non-abelian anomalies for the $SU(\kappa^{(a)}_j)$ factors forbids this. Indeed, the authors of \cite{Hanany:2018hlz} obtain the following constraint on the ranks of neighboring nodes:
\be
\kappa^{(a)}_{j} - \frac{1}{2}\left(\kappa^{(a+1)}_{j}+\kappa^{(a-1)}_{j}+\kappa^{(a)}_{j+1}+\kappa^{(a)}_{j-1}\right)+ \frac{1}{4}\left(\kappa^{(a+1)}_{j+1}+\kappa^{(a-1)}_{j+1}+\kappa^{(a+1)}_{j-1}+\kappa^{(a-1)}_{j-1}\right)=0,
\ee
which we can rewrite as:
\be
\sum_{b=1}^{r-1}\sum_{k=0}^{n-1} C^{{A}_{r-1}}_{ab}C^{\widehat{A}_{n-1}}_{jk} \kappa^{(b)}_{k} =0\qquad a=1,\dots,r-1,\qquad j=0,\dots n-1.
\ee
Since the Cartan matrix $C^{{A}_{r-1}}_{ab}$ is positive definite one finds $\sum_{k=0}^{n-1}C^{\widehat{A}_{n-1}}_{jk} \kappa^{(a)}_{k}=0$, so $\kappa^{(a)}_j$ is proportional to the imaginary root of $\widehat{A}_{n-1}$. That is:
\be
\vec{\kappa}^{(a)} = \kappa^{(a)}(1,\dots,1),
\ee
and we find such more general D3 branes configurations are not allowed in the M-string SCFT\footnote{ More general configurations of D3 branes can occur, on the other hand, in the case of the 6d M-string orbifold SCFTs \cite{wip}.}.\newline

\begin{figure}
\begin{center}
\includegraphics[width=0.5\textwidth]{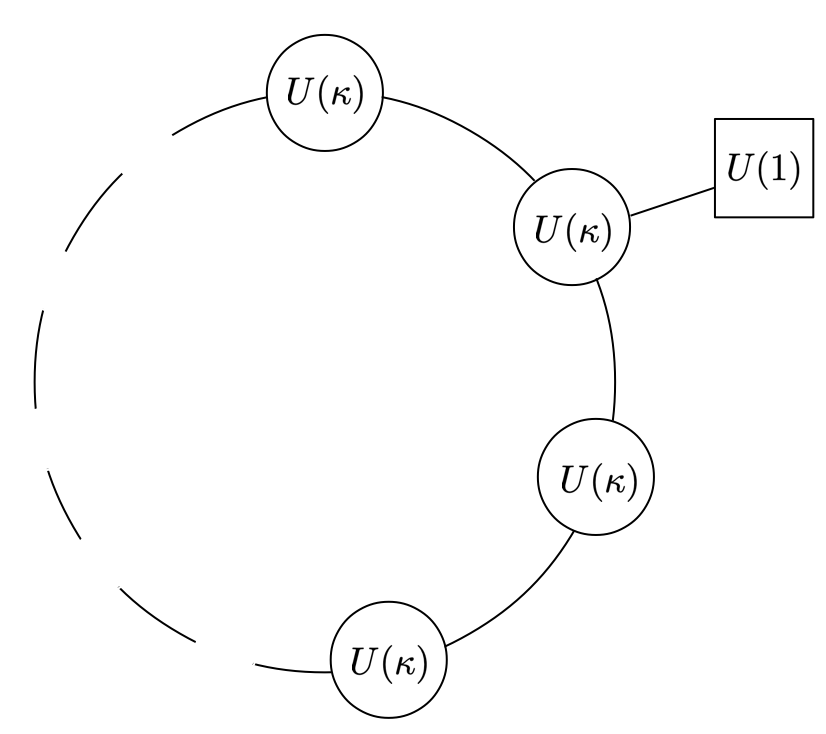}
\caption{The 3d Kronheimer-Nakajima quiver $\mathcal{KN}_{n,\kappa}$ whose Coulomb branch corresponds to the moduli space of $\kappa$ U(1) instantons on $\mathbb{C}^2/\mathbb{Z}_n$. In this quiver, hypermultiplets are represented by lines connecting gauge and global symmetry nodes.}
\label{fig:3KN}
\end{center}
\end{figure}

It is also worth to mention another interpretation of the $\mathcal{N}=(0,4)$ quivers we obtained, which clarifies their interpretation as describing the degrees of freedom of instanton strings. Namely, we can start by considering the following brane configuration in Type IIB string theory with the $r$ vertical NS5 branes removed:
\be
\label{tab:IIBKN}
\begin{tabular}{r|cccccccccc}
  & 0 & 1 & 2 & 3 & 4 & $\tilde{5}$ & 6& 7 & 8 & 9 \\
\hline
$n$ NS5 & $\times$ & $\times$ & $\times$ & $\times$ & $\times$ & & $\times$ &  &  & \\
D5 & $\times$ & $\times$ & $\times$ &  $\times$& $\times$ &  & $\times$ &  &  & \\   
\hline
$\boldsymbol{\kappa}$ D3 & $\times$ & $\times$ &  &  & &$\times$  &$\times$ &   && \\
\end{tabular}
\ee
This configuration gives rise to the 3d $\mathcal{N}=4$ Kronheimer-Nakajima \cite{kronheimer1990yang} quiver gauge theory $\mathcal{KN}_{n,\kappa^{(a)}}$ \cite{Hanany_1997} of figure \ref{fig:3KN},
 whose Coulomb branch is given by the moduli space of $\kappa^{(a)}$ $U(1)$ instantons on $\mathbf{ALE}_n$. Placing NS5 branes along directions $x_0,\dots,\tilde{x}_5$, then, imposes $(0,4)$ boundary conditions 
 \cite{Chung:2016pgt,Hanany:2018hlz} (see also \cite{Gaiotto:2016wcv,Costello:2018fnz}) for the 3d $\mathcal{N}=4$ multiplets, and gives rise to an interface between KN quiver gauge theories. We denote this interface by $\mathbf{NS}_n$, which as we have seen above involves a $\mathfrak{su}(n)_1$ current algebra is sensitive to a choice of monodromy $\omega^{KK}\in\mathbb{Z}_n$. Schematically, then, we can construct the 2d quiver gauge theory $\QQ$ by stacking the following collection of Kronheimer-Nakajima quivers along direction $x_6$:
\bea
\resizebox{\textwidth}{!}{
$
\mathcal{KN}_{n,0}
\boldsymbol{\bigg{]}}
\mathbf{NS}_n
\boldsymbol{\bigg{[}}
\mathcal{KN}_{n,\kappa^{(1)}}
\boldsymbol{\bigg{]}}
\mathbf{NS}_n
\boldsymbol{\bigg{[}}
\mathcal{KN}_{n,\kappa^{(2)}}
\boldsymbol{\bigg{]}}
\mathbf{NS}_n
\boldsymbol{\bigg{[}}
\quad
\dots
\quad 
\boldsymbol{\bigg{]}}
\mathbf{NS}_n
\boldsymbol{\bigg{[}}
\mathcal{KN}_{n,\kappa^{(r-1)}}
\boldsymbol{\bigg{]}}
\mathbf{NS}_n
\boldsymbol{\bigg{[}}
\mathcal{KN}_{n,0}
$
}
\nonumber
\eea
\newline

\section{The partition function on $T^2\times\mathbb{C}^2/\mathbb{Z}_n$}
\label{sec:Mtogether}
We can assemble a partition function for the rank-$r$ 6d M-string SCFT by summing over over contributions from configurations with different numbers of \hyphenation{wrap-ped} wrapped D3 branes, corresponding to different instanton number\footnote{ We leave out the classical terms and focus on the perturbative and instanton contributions.}:
\bea
\nonumber
&&\hskip-0.75in
\mathcal{Z}^{\boldsymbol{\omega}^{KK}}_{T^2\times\mathbb{C}^2/\mathbb{Z}_n}(\vec{\xi},\boldsymbol{\varphi},\epsilon_+,\epsilon_-,m,\tau)
=
q^{\frac{n\,r}{24}}
\times
\mathcal{Z}^{\text{BPS particles}}_{T^2\times\mathbb{C}^2/\mathbb{Z}_n}(\epsilon_+,\epsilon_-,m,\tau)
\\
&\times&
\sum_{
\boldsymbol{\kappa}=(\kappa^{(1)},\dots,\kappa^{(r-1)})\in\mathbb{Z}^{r-1}_{\geq 0}
}
e^{-\sum_{a=1}^{r-1} \kappa^{(a)}\varphi^{(a)}} \mathbb{E}^{\boldsymbol{\omega}^{KK}}_{\kappa}(\vec{\xi},\epsilon_+,\epsilon_-,m,\tau)
\label{eq:zfull}
\eea
where  $q=e^{2\pi i \tau}$. The partition function factorizes into contributions from BPS particles, which appear on the first line, and contributions from M-strings which appear on the second. The partition function depends on the chemical potentials $\boldsymbol{\varphi}$ conjugate to the instanton charge of the strings, which coincide with the separations $a^{(a+1)}-a^{(a)}$ between neighboring NS5 branes. It also depends on chemical potentials $\vec{\xi}$ which are conjugate to the $\vec{u}^{KK,(a)}$. The $\mathfrak{su}(n)_1$ current algebra couples to the two-form fields on the NS5 branes, and as a consequence the BPS strings which are charged under the two-form fields depend on the choice of monodromies $\boldsymbol{\omega}^{KK}$ as well as on $\vec{\xi}$. On the other hand, 6d BPS particles are neutral with respect to the two-form fields and therefore are insensitive to the choice of monodromy and independent of $\vec\xi$. We first discuss the BPS particle contributions, and then turn to the contribution from the M-strings.

\subsection{The BPS particle contribution}
\label{sec:bpspc}
The contribution from BPS particles can be obtained straightforwardly by orbifold projection. On $T^2\times\mathbb{C}^2$, their contribution to the partition function is given by \cite{Iqbal:2008ra,Haghighat:2013gba}
\bea
\nonumber
&&
\!\!\!\!
\mathcal{Z}^{\text{BPS particles}}_{T^2\times\mathbb{C}^2}(\epsilon_+,\epsilon_-,\tau,m)
=\\
\nonumber
&&
\!\!\!\!
\left(\prod_{k=0}^{\infty}
Z^{BPS}_{S^1\times{\mathbb{C}^2}}(k\tau,\frac{1}{2})
Z^{BPS}_{S^1\times{\mathbb{C}^2}}((k\!+\!1)\tau,-\frac{1}{2})
Z^{BPS}_{S^1\times\mathbb{C}^2}(k\tau+m,0)
Z^{BPS}_{S^1\times\mathbb{C}^2}((k\!+\!1)\tau-m,0)\right)^r\!\!,\\
\label{eq:bpsppp}
\eea
which also coincides with the BPS partition function of 5d $\mathcal{N}=1^*\,$ $U(1)^r$ SYM that arises from the NS5 branes upon compactification, upon identifying $\mu = m-\epsilon_+$ with the mass of the adjoint hypermultiplet.
The contribution from an individual BPS state of mass $m$, which is neutral under $U(1)_x$ and carries charge $k_t \in\frac{1}{2}\mathbb{Z}$ under the Cartan of  $SU(2)_t$, is given by:
\bea
\nonumber
Z^{BPS}_{S^1\times\mathbb{C}^2}(m,k_t)
&=&
\prod_{i,j=0}^\infty
(1-Q_m t^{i-j+2k_t}x^{i+j+1})^{(-1)^{2k_t}}
\\
&=&
PE\left[Q_m t^{2 k_t}\frac{t}{(1-tx)(1-t x^{-1})}\right]^{(-1)^{2k_t}},
\eea
where $Q_m = e^{2\pi i m}=e^{2\pi i \mu}t$  and $ PE[f(x_1,\dots,x_k)] = \prod_{n=1}^\infty \exp\left(\frac{1}{n}f(x_1^n,\dots,x_n^n)\right)$ is the plethystic exponential. Using this, one can write:
\be
\mathcal{Z}^{\text{BPS particles}}_{T^2\times\mathbb{C}^2}(\epsilon_+,\epsilon_-,\tau,m)
\!
=
\!
PE\left[\!
\left(\!Q_m\!-\!t\!+\!\frac{q(1-\frac{Q_m}{t})(1- Q_m t)}{(1-q)Q_m}\right)\frac{t^{} }{(1-t x)(1-\frac{t}{x})}\!\right]^r
\!
.
\ee

The BPS particles' partition function receives contributions from operators with different $SU(2)_x\times SU(2)_t$ charge. To perform the orbifold projection, recalling that $\mathbb{Z}_n$ is embedded into $SU(2)_x$ we simply keep the states whose charge under the Cartan of $SU(2)_x$ is a multiple of $n$, and throw out the rest. In other words, we have the following:
\be
Z^{BPS}_{S^1\times\mathbb{C}^2/\mathbb{Z}_n}(m,k_t)
=
\prod_{\substack{i,j=0\\i+j+1=0 \text{ mod }n}}^\infty
(1-e^{2\pi i m} t^{i-j+2k_t}x^{i+j+1})^{(-1)^{2k_t}}.
\ee
Then, the BPS particles' contributions to the partition function on $ T^2\times\mathbb{C}^2/\mathbb{Z}_n$, $\mathcal{Z}^{\text{BPS particles}}_{T^2\times\mathbb{C}^2/\mathbb{Z}_n}$, is obtained by replacing
\be
Z^{BPS}_{S^1\times\mathbb{C}^2} \to Z^{BPS}_{S^1\times\mathbb{C}^2/\mathbb{Z}_n}
\ee
everywhere on the right hand side of equation \eqref{eq:bpsppp}. In terms of the plethystic exponential, one can write:
\bea
\nonumber
&&
\mathcal{Z}^{\text{BPS particles}}_{T^2\times\mathbb{C}^2/\mathbb{Z}_n}(\epsilon_+,\epsilon_-,\tau,m)
\\
\nonumber
&=&
PE\left[r\left(Q_m-t+\frac{q}{1-q}\frac{(1-\frac{Q_m}{t})(1- Q_m t)}{Q_m}\right)\sum_{j=0}^{n-1}\frac{t^{} }{n(1-e^{\frac{2\pi i j}{n}}t x)(1-e^{-\frac{2\pi i j}{n}}\frac{t}{x})}\right]
\label{eq:bpsc2zn}\\
&=&
PE\left[r\left(Q_m-t+\frac{q}{1-q}\frac{(1-\frac{Q_m}{t})(1- Q_m t)}{Q_m}\right)t\,\mathcal{H}_{\mathbb{C}^2/\mathbb{Z}_n}\right],
\label{eq:bpspleth}
\eea
where the last factor in the plethystic exponential is be the Hilbert series of $\mathbb{C}_2/\mathbb{Z}_n$:
\be
\mathcal{H}_{\mathbb{C}^2/\mathbb{Z}_n} = \frac{1-t^{2n}}{1-t^2}\frac{1}{(1-t^n x^n)(1-t^nx^{-n})}.
\ee

\subsection{M-string contributions}
\label{sec:bpssc}
Let us now turn to the contributions from the M-strings, which are the focus of this paper. The contribution of a bound state of $\boldsymbol{\kappa} = (\kappa^{(1)},\dots,\kappa^{(r-1)})$ M-strings is weighed by an overall factor of $e^{-\sum_{a=1}^{r-1}\kappa^{(a)}\varphi^{(a)}}$, where $\sum_{a=1}^{r-1}\kappa^{(a)}\varphi^{(a)}$ is the tension of the bound state of strings. The bound state's contribution to the partition function is given in terms of the elliptic genus $\mathbb{E}^{\boldsymbol{\omega}^{KK}}_{\kappa}$ of the 2d $\mathcal{N}=(0,4)$ theory $\QQ$ evaluated for the choice of monodromies $\boldsymbol{\omega}^{KK}$, which is defined as follows:
\be
\resizebox{\textwidth}{!}{
$
\mathbb{E}^{\boldsymbol{\omega}^{KK}}_{\boldsymbol{\kappa}}(\vec{\xi},\epsilon_+,\epsilon_-,m,\tau)
=
\Tr (-1)^{F} q^{H_L}  \overline{q}^{H_R}e^{2\pi i \epsilon_+(J_R+J_t)}e^{2\pi i \epsilon_- J_x}
e^{2\pi i m J_m}
e^{2\pi i \vec{\zeta}\cdot\vec{J}_{\mathfrak{su}(n)}}.
$
}
\ee
where $\vec{\xi}=C^{A_{n-1}}\cdot \vec{\zeta} $, and $J_R,$ $J_m$, $J_x,$ and $J_t$ are respectively the generators for the $SU(2)_R$ R-symmetry, for $U(1)^{diag}$, for $U(1)_x$ and for the Cartan of $SU(2)_t$. On the other hand, $\vec{J}_{\mathfrak{su}(n)}$  are the generators for the cartan of the $\mathfrak{su}(n)_1$ current algebra which appears as a global symmetry of $\QQ$. The elliptic genus can be computed by localization \cite{Benini:2013xpa} and is given by the following integral over the holonomies $z^{(a)}_{j,1},\dots,z^{(a)}_{j,k}$  along $T^2$ for the gauge groups $G^{(a)}_j=U(\kappa^{(a)}_j)$:
\bea
\nonumber
\mathbb{E}^{\boldsymbol{\omega}^{KK}}_{\boldsymbol{\kappa}}
\hskip-.05in
=
\hskip-.05in
\bigg(
\prod_{a=1}^{r-1}\frac{1}{\kappa^{(a)}!}
\bigg)^n
\hskip-.1in
\int &&\!\!\!\!\!\!\!\!\!
\left[
\prod_{j=0}^{n-1}
\bigg(\prod_{a=1}^{r-1} Z_{V^{(a)}_j} Z_{X^{(a)}_j}\bigg)
\hskip-.05in
\bigg(\prod_{a=1}^{r-2} Z_{Y^{(a)}_j,\Psi^{(a)}_j,\widetilde{\Psi}^{(a)}_j}\bigg)\right]
\hskip-.05in
\bigg(\prod_{a=1}^{r-1}Z_{W^{(a)},\Sigma^{(a)},\Theta^{(a)}}\bigg)\\
&\times&
\bigg(\prod_{a=1}^{r}Z_{\mathcal{H}}Z^{\omega^{KK,(a)}}_{\mathfrak{su}(n)^{(a)}}\bigg),
\label{eq:ellint}
\eea
which is evaluated by summing over Jeffrey-Kirwan residues. The first row of equation \eqref{eq:ellint} captures the contributions from fields that arise from the degrees of freedom of the D3 branes. These are given by: 
\bea
\nonumber
Z_{V^{(a)}_j}
&=&
\left(\prod_{k=1}^{\kappa^{(a)}}\frac{\eta^2 dz^{(a)}_{j,k}}{2\pi i}\frac{\theta_1(2\epsilon_+)}{\eta}\right)
\left(\prod_{\substack{k,l=1 \\ k\neq l}}^{\kappa^{(a)}}\frac{\theta_1(z^{(a)}_{j,k}-z^{(a)}_{j,l})\theta_1(2\epsilon_+ + z^{(a)}_{j,k}-z^{(a)}_{j,l})}{\eta^2}\right);\\
\nonumber
Z_{X^{(a)}_j}
&=&
\left(\prod_{k,l=1}^{\kappa^{(a)}}\frac{\eta^2}{\theta_1(\epsilon_++\epsilon_-+ z^{(a)}_{j,k}-z^{(a)}_{j+1,l})\theta_1(\epsilon_+-\epsilon_-- z^{(a)}_{j,k}+z^{(a)}_{j+1,l})}\right);\\
\nonumber
Z_{W^{(a)},\Sigma^{(a)},\Theta^{(a)}}
&=&
\left(\prod_{k=1}^{\kappa^{(a)}}\frac{\theta_1(z^{(a)}_{0,k}-s^{(a-1)})\theta_1(-z^{(a)}_{0,k}+s^{(a+1)})}{\theta_1(\epsilon_++ z^{(a)}_{0,k}-s^{(a)})\theta_1(\epsilon_++s^{(a)}-z^{(a)}_{0,k})}\right);\\
\nonumber
Z_{Y^{(a)}_j,\Psi^{(a)}_j,\widetilde{\Psi}^{(a)}_j}
&=&
\!
\left(
\prod_{k=1}^{\kappa^{(a)}}\prod_{l=1}^{\kappa^{(a+1)}}
\frac{\theta_1(\epsilon_-\!+\! z^{(a)}_{j,k}\!-\!z^{(a+1)}_{j+1,l})\theta_1(\epsilon_- \!-\! z^{(a)}_{j,k}\!+\!z^{(a+1)}_{j-1,l})}
{\theta_1(\epsilon_+\!+\! z^{(a)}_{j,k}\!-\!z^{(a+1)}_{j,l})\theta_1(\epsilon_+-\!z^{(a)}_{j,k} \!+\! z^{(a+1)}_{j,l})}
\right)
\!,
\\
\eea
where
\bea
\eta
&=&
\eta(\tau) = q^{\frac{1}{24}}\prod_{k=1}^\infty(1-q^k),\\
\theta_1(z)
&=&
\theta_1(z,\tau) = i q^{\frac{1}{8}}e^{-\pi i z}\prod_{k=1}^\infty(1-q^k)(1-q^{k-1}e^{2\pi i z})(1-q^{k}e^{-2\pi i z})
\eea
and the $s^{(a)}$ denote the holonomies of $U(1)^{(a)}$, which due to the Stuckelberg mechanism satisfy the constraint $s^{(a)}=s^{(a-1)}+m$. The second row of equation \eqref{eq:ellint} captures the contributions that arise from coupling to the $\mathfrak{su}(n)_1$ current algebra, as well as a decoupled contribution from $r$ copies of the Heisenberg algebra. The latter is given simply by:
\be
Z_{\mathcal{H}}^r = \eta(\tau)^{-r},
\ee
while the former is given by:
\be
Z^{\omega^{KK,(a)}}_{\mathfrak{su}(n)^{(a)}} = \widehat{\chi}^{\mathfrak{su}(n)_1}_{\omega^{KK,(a)}}(\vec{\xi}^{(a)},\tau),
\ee
where 
\be
\xi^{(a)}_j=\xi_j+(C^{\widehat{A}_{n-1}}\cdot(\vec{Z}^{(a)}-\vec{Z}^{(a-1)}))_j
\ee
for $j=1,\dots,n-1$,
\be
Z_j^{(a)}= \sum_{k=1}^{\kappa^{(a)}}z_{j,k}^{(a)}
\ee
is the combination of chemical potentials associated to the determinant representation of $G^{(a)}_j= U(\kappa^{(a)})$, and the level 1 $\mathfrak{su}(n)$ character corresponding to the integrable highest weight module $V_\lambda$ associated to the Dynkin label $\lambda^\omega$ with entries $\lambda^\omega_j = \delta_{j,\omega}$ for $\omega\in\mathbb{Z}_n$ is given by:
\bea
\nonumber
\widehat{\chi}^{\mathfrak{su}(n)_1}_\omega(\vec{\xi},\tau)
&=&
\Tr_{V_{\lambda^\omega}}\left(q^{H_L}e^{2\pi i \vec{\xi}\cdot (C^{A_{n-1}})^{-1}\cdot\vec{J}_{\mathfrak{su}(n)}}\right)
\\
\nonumber
&=&
\eta(\tau)^{-n+1}
\hspace{-.4in}
\sum_{\substack{\vec{u}\in \mathbb{Z}^{n-1}\\\sum_{j=1}^{n-1}ju_j =\omega\mod n}}
\hspace{-.4in}
q^{\frac{1}{2}\vec{u} \cdot (C^{A_{n-1}})^{-1}\cdot \vec{u}}
e^{2\pi i \vec\xi\cdot(C^{A_{n-1}})^{-1}\cdot\vec{u}}.\\
\label{eq:sunchar}
\eea
The holonomies $Z_j^{(0)}$ and $Z_j^{(r)}$ are to be set to $0$ since there are no D3 branes supported to the left (respectively right) of the first (resp. $r$-th) NS5 brane.

 It is convenient to shift the integration variables $z^{(a)}_{j,k}$ as follows:
\be
z^{(a)}_{j,k}\to z^{(a)}_{j,k}+ s^{(a)},
\ee
under which the one-loop factors $Z_{V^{(a)}_j}$ and $Z_{X^{(a)}_j}$ as well as $Z^{\omega^{KK,(a)}}_{\mathfrak{su}(n)^{(a)}}$ are invariant, $Z_{W^{(a)},\Sigma^{(a)},\Theta^{(a)}}$ and $Z_{Y^{(a)}_j,\Psi^{(a)}_j,\widetilde{\Psi}^{(a)}_j}$ become:
\bea
\nonumber
Z_{W^{(a)},\Sigma^{(a)},\Theta^{(a)}}
&=&
\left(\prod_{k=1}^{\kappa^{(a)}}\frac{\theta_1(m+z^{(a)}_{0,k})\theta_1(m-z^{(a)}_{0,k})}{\theta_1(\epsilon_++ z^{(a)}_{0,k})\theta_1(\epsilon_+-z^{(a)}_{0,k})}\right);\\
\nonumber
Z_{Y^{(a)}_j,\Psi^{(a)}_j,\widetilde{\Psi}^{(a)}_j}
&=&
\!
\left(
\prod_{k=1}^{\kappa^{(a)}}\prod_{l=1}^{\kappa^{(a+1)}}
\frac{\theta_1(\epsilon_-\! -\! m\!+\! z^{(a)}_{j,k}\!-\!z^{(a+1)}_{j+1,l})\theta_1(\epsilon_- \!+\! m\!-\! z^{(a)}_{j,k}\!+\!z^{(a+1)}_{j-1,l})}
{\theta_1(\epsilon_+ \!-\! m\!+\! z^{(a)}_{j,k}\!-\!z^{(a+1)}_{j,l})\theta_1(\epsilon_+ \!+\! m\!-\!z^{(a)}_{j,k} \!+\! z^{(a+1)}_{j,l})}
\right).
\\
\eea

In fact, it is possible to give a combinatorial formula for the sum over residues in equation \eqref{eq:ellint}, building upon \cite{Dey:2013fea} as well as \cite{Gadde:2015tra}. Denote by $\mathcal{Y}_{\kappa^{(a)}}$ the set of Young diagrams $\mathbf{Y}^{(a)}$ of size $ n \kappa^{(a)}$. Denote by $(\mathfrak{b}^1,\mathfrak{b}^2)$ the coordinates of a box inside a Young diagram. To a given box $\mathfrak{b} = (\mathfrak{b}^1,\mathfrak{b}^2)$, we assign the integer 
\be
\omega_\mathfrak{b} = -\mathfrak{b}^1+\mathfrak{b}^2 \qquad \text{mod } n.
\ee
 The residues associated to the gauge holonomies $\vec{z}^{(a)}$ are labeled by elements of a subset $\widetilde{\mathcal{Y}}_{\kappa^{(a)}} $ of $\mathcal{Y}_{\kappa^{(a)}}$. The elements of $\widetilde{\mathcal{Y}}_{\kappa^{(a)}} $ are Young diagrams that contain exactly $\kappa^{(a)}$ boxes $\mathfrak{b}$ such that $\omega_\mathfrak{b} = j$, for all $j=0,\dots, n-1$. Residues are prescribed by making the assignment 
\be
z^{(a)}_{j,k}\to \epsilon_++(\epsilon_++\epsilon_-)\mathfrak{b}_{j,k}^{(a),1}+(\epsilon_+-\epsilon_-)\mathfrak{b}_{j,k}^{(a),2}
\label{eq:zass}
\ee
for all gauge holonomies, where $\{\mathfrak{b}_{j,k}^{(a)}\}_{k=1,\dots,\kappa^{(a)}}$ is the set of boxes for which $\omega_{\mathfrak{b}}=j$. There are $\prod_{a=1}^{r-1}(\kappa^{(a)}!)^n$ equivalent ways to make this assignment, which compensates for the combinatorial factor appearing in front of equation \eqref{eq:ellint}.

The elliptic genus is obtained by summing over the residues:
\bea
\nonumber
\mathbb{E}^{\boldsymbol{\omega}^{KK}}_{\boldsymbol{\kappa}}
\hskip-.05in
&=&
\bigg(
\prod_{a=1}^{r-1}\frac{1}{\kappa^{(a)}!}
\bigg)^n
\hskip-.17in
\sum_{\substack{Y^{(1)}\in \widetilde{\mathcal{Y}}_{\kappa^{(1)}}\\\dots\\Y^{(r)}\in \widetilde{\mathcal{Y}}_{\kappa^{(r-1)}}}}
\hskip-.17in
\left(\frac{\theta_1(2\epsilon_+)}{\eta}\right)^{n\sum_{a=1}^{r-1}\kappa^{(a)}}
\hskip-.1in
\bigg(\prod_{j=0}^{n-1}\prod_{a=1}^{r-1} Z_{X^{(a)}_j}\bigg)
\bigg(\prod_{j=0}^{n-1}\prod_{a=1}^{r-2} Z_{Y^{(a)}_j,\Psi^{(a)}_j,\widetilde{\Psi}^{(a)}_j}\bigg)
\\
&\times&
\bigg(\prod_{a=1}^{r-1}Z_{W^{(a)},\Sigma^{(a)},\Theta^{(a)}}\bigg)
\bigg(\prod_{a=1}^{r}Z_{\mathcal{H}}Z^{\omega^{KK,(a)}}_{\mathfrak{su}(n)^{(a)}}\bigg)
\bigg]
\bigg\vert_{z^{(a)}_{j,k}\to \epsilon_++(\epsilon_++\epsilon_-)\mathfrak{b}_{j,k}^{(a),1}+(\epsilon_+-\epsilon_-)\mathfrak{b}_{j,k}^{(a),2}},
\label{eq:ellipticgenus}
\eea
where  the notation $\prod \!\phantom{\vert}^{'}$ signifies that any occurrence of $\theta_1(0)$ in the product is to be replaced by a factor of $\eta$. In section \ref{sec:examples} we will present explicit expressions for the elliptic genus in several different examples of theories $\QQ$.

\subsection{Comments on the partition function}
\label{sec:comm}
In this section we remark on some salient features of the partition function \eqref{eq:zfull}.\\

\paragraph{Dependence on the spacetime topology.} The main new feature that arises on the background $T^2\times\mathbb{C}^2/\mathbb{Z}_n$ is that the partition function is no longer uniquely defined, but depends on the choice of discrete data $\boldsymbol{\omega}^{KK}$. Expanding what we have explained in the introduction this dependence has a twofold effect. On one hand we expect the in general the partition vectors of 6d SCFTs with a non-trivial defect group \cite{DelZotto:2015isa} get promoted to \textit{partition matrices}: on top of the dependence on the discrete data from the bulk 7d Chern Simons theory that specifies the global form of the gauge group (as explained in \cite{Tachikawa:2013hya}), the partition vector for these models receive further choices of discrete data from the actual spacetime boundary, corresponding to the 6d uplift of the prescription of a flat connection at infinity for the 5d $\mathcal N=1^*$ $U(1)^r$ theory.\footnote{\ We believe these data should be more precisely encoded in a choice of ``flat 2-gerbe'' on the Lens space $S^3/\mathbb Z_n$.}

At the level of the solitonic degrees of freedom, we see that the extra data giving rise to a partition matrix has the effect, as discussed in section \ref{sec:bpsdof} above, to also give rise to a relativization of the worldsheet theories of BPS strings, which for the ALE background we consider become bulk-boundary systems with a boundary capable of detecting the prescription of flat connection at infinity.

\paragraph{$\mathcal{N}=(2,0)$ limit.} Notice that, the theory $\QQ$ is nontrivial even in the zero-string sector, i.e. it is given by the Heisenberg and current algebra contributions. At the level of elliptic genus,
\be
\mathbb{E}^{\boldsymbol{\omega}^{KK}}_{\boldsymbol{0}} = \eta(\tau)^{-r}\prod_{a=1}^r\widehat{\chi}^{\mathfrak{su}(n)_1}_{\omega^{KK,(a)}}(\vec\xi,\tau).
\label{eq:e0}
\ee

It is interesting to note that, as in the $T^2\times\mathbb{C}^2$ case, we can recover the partition function of the $\mathcal{N}=(2,0)$ theory on $T^2\times\mathbb{C}^2/\mathbb{Z}_k$ by specializing
\be
m\to \epsilon_+.
\ee
In this limit, it is straightforward to see that the elliptic genus for any bound state of M-strings is zero due to numerator terms in $Z_{W^{(a)},\Sigma^{(a)},\Theta^{(a)}}$, which vanish upon setting the integration variables to \eqref{eq:zass}. As a consequence, the sum over BPS string sectors in the partition function \eqref{eq:zfull} collapses to just the leading order term \eqref{eq:e0} corresponding to zero string charge. At the same time, it is trivial to see that in the same limit the product over BPS particle contributions becomes identically 1. In other words, we find that
\bea
\mathcal{Z}^{\boldsymbol{\omega}^{KK}}_{T^2\times\mathbb{C}^2/\mathbb{Z}_n}(\vec{\xi},\boldsymbol{\varphi},\epsilon_+,\epsilon_-,\epsilon_+,\tau)
=
q^{\frac{n\,r}{24}}
\eta(\tau)^{-r}\prod_{a=1}^r\widehat{\chi}^{\mathfrak{su}(n)_1}_{\omega^{KK,(a)}}(\vec\xi,\tau).
\eea
This agrees with the equivariant partition function  on $\mathbb{C}^2/\mathbb{Z}_n$ of maximally supersymmetric $U(r)$ SYM on the Coulomb branch \cite{Bruzzo:2013daa}, which is simply given by $r$ copies of the the Vafa-Witten partition function \cite{Vafa:1994tf} for the maximally supersymmetric $U(1)$ SYM on $\mathbb{C}^2/\mathbb{Z}_n$ (see also \cite{Fujii:2005dk}).

\paragraph{Nekrasov master formula.} Building from remarks in \cite{DelZotto:2021gzy} we are lead to conjecture that the partition function \eqref{eq:zfull} is also invariant under the Kahler deformations of $\mathbb{C}^2/\mathbb{Z}_n$. In particular, the expectation is that
\be
\mathcal{Z}^{\boldsymbol{\omega}^{KK}}_{T^2\times\mathbb{C}^2/\mathbb{Z}_n}(\vec{\xi},\boldsymbol{\varphi},\epsilon_+,\epsilon_-,m,\tau)
=
\mathcal{Z}^{\boldsymbol{\omega}^{KK}}_{T^2\times\widetilde{\mathbb{C}^2/\mathbb{Z}_n}}(\vec{\xi},\boldsymbol{\varphi},\epsilon_+,\epsilon_-,m,\tau),
\ee
where $\mathcal{Z}^{\text{BPS particles}}_{T^2\times\widetilde{\mathbb{C}^2/\mathbb{Z}_n}}$ is the partition function on the resolution of the $\mathbf{ALE}_n$ singularity. As reviewed in section \ref{sec:Mstring}, the M-string SCFT on a circle admits a dual interpretation as a 5d $\mathcal{N}=1^*$ $U(r)$ gauge theory, so one expects the partition function \eqref{eq:zfull} to coincide with the partition function of the $\mathcal{N}=1^*$ $U(r)$ gauge theory on $\widetilde{\mathbb{C}^2/\mathbb{Z}_n}$. The latter can be inferred from the 4d computations in \cite{Bruzzo:2013daa} (see also \cite{Fucito:2004ry,Fucito:2006kn,Bonelli:2011jx,Bonelli:2012ny,Ito:2013kpa,Alfimov:2013cqa,Dey:2013fea,Mekareeya:2015bla} for related results), and indeed as detailed in appendix \ref{sec:compar} we find agreement between the two in the case $r=n=2$ (upon making some small adaptations to the 5d gauge theoretic partition function). Moreover, the relation to gauge theory on $\mathbb{C}^2/\mathbb{Z}_n$ suggests an alternative expression for the partition function based on the Nekrasov master formula \cite{Nekrasov:2003vi,Gasparim:2009sns,Bonelli:2012ny,Bruzzo:2013daa}, which factorizes into contributions localized at the fixed points of the equivariant action on $\widetilde{\mathbb{C}^2/\mathbb{Z}_n}$. Applied to the M-string SCFT, it predicts:
\bea
\nonumber
&&
\hspace{-.25in}
\mathcal{Z}^{\boldsymbol{\omega}^{KK}}_{T^2\times\mathbb{C}^2/\mathbb{Z}_n}(\vec{\xi},\boldsymbol{\varphi},\epsilon_+,\epsilon_-,m,\tau) =
\hspace{-.58in}
\sum_{\substack{\vec{u}^{KK,(a)}\in \mathbb{Z}^{n-1}\\\sum_{j=1}^{n-1}j\vec{u}^{KK,(a)}_j =\omega^{KK,(a)}\mod n}}
\hspace{-.58in}
q^{\frac{1}{2}\vec{u}^{KK,(a)} \cdot (C^{A_{n-1}})^{-1}\cdot \vec{u}^{KK,(a)}}
e^{2\pi i \vec\xi\cdot(C^{A_{n-1}})^{-1}\cdot\vec{u}^{KK,(a)}}
\\
&\times&
\prod_{j=1}^n
\mathcal{Z}_{T^2\times\mathbb{C}^2}(\vec{\xi},\boldsymbol{\varphi}_{\vec{u}^{KK,(a)}}[j],\epsilon_+[j],\epsilon_-[j],m,\tau).
\eea
Here,
\bea
\nonumber
&&\hskip-0.75in
\mathcal{Z}_{T^2\times\mathbb{C}^2}(\boldsymbol{\varphi},\epsilon_+,\epsilon_-,m,\tau)
=
q^{\frac{r}{24}}
\times
\mathcal{Z}^{\text{BPS particles}}_{T^2\times\mathbb{C}^2}(\epsilon_+,\epsilon_-,m,\tau)
\\
&
\times
&
\sum_{
\boldsymbol{\kappa}=(\kappa^{(1)},\dots,\kappa^{(r-1)})\in\mathbb{Z}^{r-1}_{\geq 0}
}
e^{-\sum_{a=1}^{r-1} \kappa^{(a)}\varphi^{(a)}} \mathbb{E}_{\boldsymbol{\kappa}}(\epsilon_+,\epsilon_-,m,\tau)
\label{eq:zfullc2}
\eea
is the M-string SCFT partition function on $\mathbb{C}^2$ (which coincides to the $n=1$ case in equation \eqref{eq:zfull}), and
\bea
\varphi_{\vec{u}^{KK,(a)}}^{(a)}[j] &=& \pi i -2\pi i({a}_{\vec{u}^{KK,(a)}}^{(a+1)}[j]-a_{\vec{u}^{KK,(a)}}^{(a)}[j]),\,\,\,\\
a_{\vec{u}^{KK,(a)}}^{(a)}[j] &=& a^{(a)}\!+\!((C^{A_{n-1}})^{-1}\!\cdot\! \vec{u}^{KK,(a)})_j\epsilon_1[j]\!+\!((C^{A_{n-1}})^{-1}\!\cdot\! \vec{u}^{KK,(a)})_{j-1}\epsilon_2[j],\,\,\,
\eea
where we defined
\bea
\epsilon_1[j]&=&\epsilon_+[j]+\epsilon_-[j],\\
 \epsilon_2[j]&=&\epsilon_+[j]-\epsilon_-[j].
\eea
This gives rise to a particularly simple prediction for the partition function:
\bea
\nonumber
&&\mathcal{Z}^{\boldsymbol{\omega}^{KK}}_{T^2\times\mathbb{C}^2/\mathbb{Z}_2}(\vec{\xi},\boldsymbol{\varphi},\epsilon_+,\epsilon_-,m,\tau)
=
q^{\frac{rn}{24}}
\prod_{j=1}^n
\mathcal{Z}^{\text{BPS particles}}_{T^2\times\mathbb{C}^2}(\epsilon_+[j],\epsilon_-[j],m,\tau)
\\
\nonumber
&&
\times
\hspace{-.3in}
\sum_{
\boldsymbol{\kappa}=(\kappa^{(1)},\dots,\kappa^{(r-1)})\in\mathbb{Z}^{r-1}_{\geq 0}
}
\hspace{-.3in}
e^{-\sum_{a=1}^{r-1} \kappa^{(a)}\varphi^{(a)}}
\sum_{\vec{\boldsymbol\kappa}
\in\mathcal{K}_{\boldsymbol\kappa}}
\prod_{a=1}^r\widehat{\chi}^{\mathfrak{su}(n)_1}_{\omega^{{KK},(a)}}(\vec\xi_{\vec{\kappa}^{(a-1)},\vec{\kappa}^{(a)}},\tau)
\prod_{j=1}^{n} \mathbb{E}_{\boldsymbol{\kappa}_j}(\epsilon_+[j],\epsilon_-[j],m,\tau).\\
\eea
Here, the elements of the set $\mathcal{K}_{\boldsymbol{\kappa}}$ are $(r+1)$-tuples $\vec{\boldsymbol{\kappa}} = (\vec{\kappa}^{(0)},\vec{\kappa}^{(1)},\dots,\vec{\kappa}^{(r)})\in (\mathbb{Z}_{\geq 0}^n)^{r+1}$ such that $\vec{\kappa}^{(0)}=0=\vec{\kappa}^{(r)}$ and $\sum_{j=1}^n\vec{\kappa}^{(a)}=\kappa^{(a)}$ for $a=1,\dots,r-1$, and
\be
\xi_{\vec{\kappa}^{(a-1)},\vec{\kappa}^{(a)},j}
=
\xi_j
+
(\kappa_j^{(a)}-\kappa_j^{(a-1)})\epsilon_1[j]
+
(\kappa_{j+1}^{(a)}-\kappa_{j+1}^{(a-1)})\epsilon_2[j+1].
\ee
Indeed, on the one hand is straightforward to verify that
\be
\prod_{j=1}^n
\mathcal{Z}^{\text{BPS particles}}_{T^2\times\mathbb{C}^2}(\epsilon_+[j],\epsilon_-[j],m,\tau)
=
\mathcal{Z}^{\text{BPS particles}}_{T^2\times\mathbb{C}^2/\mathbb{Z}_n}(\epsilon_+,\epsilon_-,\tau,m),
\ee
where the contribution of BPS particles on $\mathbb{C}^2/\mathbb{Z}_n$ is given by equation \eqref{eq:bpspleth}. On the other hand, we get a prediction for the elliptic genus of a bound state of M-strings on $\widetilde{\mathbb{C}^2/\mathbb{Z}_n}$:
\be
\mathbb{E}^{\boldsymbol{\omega}^{KK}}_{\widetilde{\mathbb{C}^2/\mathbb{Z}_n},\boldsymbol{\kappa}}(\epsilon_+,\epsilon_-,m,\tau) = \sum_{\vec{\boldsymbol\kappa}
\in\mathcal{K}_{\boldsymbol\kappa}}
\prod_{a=1}^r\widehat{\chi}^{\mathfrak{su}(n)_1}_{\omega^{{KK},(a)}}(\vec\xi_{\vec{\kappa}^{(a-1)},\vec{\kappa}^{(a)}},\tau)
\prod_{j=1}^{n} \mathbb{E}_{\boldsymbol{\kappa}_j}(\epsilon_+[j],\epsilon_-[j],m,\tau).
\label{eq:ellreso}
\ee
It is a nontrivial fact, which we have verified explicitly in a number of examples, that this expression agrees with the elliptic genus $\mathbb{E}^{\boldsymbol{\omega}^{KK}}_{\boldsymbol{\kappa}}$ of theory $\QQ$ given by equation \eqref{eq:ellipticgenus}. In section \ref{sec:examples} we will present some checks of this relation.

\section{M-strings on $T^2\times\mathbb{C}^2/\mathbb{Z}_n$}
\label{sec:mstr}
In this section, we study some basic properties of the 2d theories $\QQ$. We start in section \ref{sec:cc} by determining their central charges, global anomalies and modular transformation properties. In section \ref{sec:examples} we study some explicit examples of elliptic genera.
\subsection{Anomalies and central charges}
\label{sec:cc}
Let us start by working out the central charges of theory $\QQ$. The leading order power of $q$ in the elliptic genus is given by $q^{\frac{2c_L-2c_R-3n\,r}{24}}$, from which one reads off the gravitational anomaly
\be
c_L - c_R = n \cdot r,
\ee
which is produced by the coupling to the $\mathcal{H}\times\mathfrak{su}(n)_1$ degrees of freedom. The right moving central charge in the IR can be read off as in \cite{Kim:2014dza} by computing 
\be
c_R = 3\Tr \gamma^3 R^2
\ee
which receives contributions from the fermions in the $(0,4)$ vector multiplets as well as in $X^{(a)}_j$ and $W^{(a)}$. The contributions from the first two cancel against each other, while from $W^{(a)}$ we find simply:
\be
c_R = 6\sum_{a=1}^{r-1}\kappa^{(a)}.
\ee

Because $\QQ$ is a relative theory, the elliptic genus transforms as a component of a vector valued, meromorphic Jacobi form. Indeed, under a modular $S$ transformation, the characters of $\mathfrak{su}(n)_1$ transform as follows:
\be
\widehat{\chi}^{\mathfrak{su}(n)_1}_j\left(\frac{\vec\xi}{\tau},-\frac{1}{\tau}\right) = e^{\frac{1}{2}\frac{2\pi i}{\tau}\vec{\xi}\cdot (C^{{A}_{n-1}})^{-1}\cdot\vec{\xi}}\sum_{k=0}^{n-1}\mathcal{S}_{jk} \widehat{\chi}_k^{\mathfrak{su}(n)_1}(\vec\xi,{\tau}),
\label{eq:sutr}
\ee
where the modular $S$-matrix has components
\be
\mathcal{S}_{jk} = e^{\frac{2\pi i}{n} j\, k} \qquad\text{ for } j,k=0,\dots,n-1.
\ee
As a consequence, the elliptic genera with different choices of monodromy $\boldsymbol{\omega}^{KK}$ transform among themselves under an S-transformation:
\bea
\nonumber
&&\mathbb{E}^{\boldsymbol{\omega}^{KK}}_{\boldsymbol{\kappa}}\left(\frac{\vec{\xi}}{\tau},\frac{\epsilon_+}{\tau},\frac{\epsilon_-}{\tau},\frac{m}{\tau},-\frac{1}{\tau}\right)
=
(-i\tau)^{-\frac{r}{2}}\times
\\
&&
\sum_{\boldsymbol{\upsilon}^{KK}=(\upsilon^{KK,(1)},\dots,\upsilon^{KK,(r)})\in \mathbb{Z}_n^r}
\hspace{-.7in}
e^{\frac{2\pi i}{\tau}f_{\boldsymbol{v}}(\vec{{\xi}},\epsilon_+,\epsilon_-,m)}
\left(\prod_{a=1}^r \mathcal{S}_{\omega^{KK,(a)}\upsilon^{KK,(a)}}\right)
\mathbb{E}^{\boldsymbol{\upsilon}^{KK}}_{\boldsymbol{v}}(\vec{\xi},\epsilon_+,\epsilon_-,m,\tau).
\eea
The overall  modular weight of $-\frac{r}{2}$ is due to the $r $ copies of the Heisenberg algebra, which are decoupled from the other degrees of freedom. The modular anomaly $f_{\boldsymbol{v}}(\vec{{\xi}},\epsilon_+,\epsilon_-,m)$ encodes 't Hooft anomalies for the global symmetries of the string worldsheet. This can be computed by using the modular transformation of the $\mathfrak{su}(n)_1$ characters (equation \eqref{eq:sutr}) and of the Jacobi theta functions: 
\be
\theta_1\left(\frac{z}{\tau},-\frac{1}{\tau}\right) = (-i)\sqrt{-i\tau}e^{\frac{2\pi i}{\tau}\frac{1}{2}z^2}\theta_1\left(z,\tau\right).
\ee
Combining the contributions from the various factors in the integrand of equation \eqref{eq:ellint}, we find that the modular anomaly is given by:
\be
f_{\boldsymbol{v}}(\vec{\xi},\epsilon_+,\epsilon_-,m)= \frac{r}{2}\, \vec{\xi} \cdot (C^{{A}_{n-1}})^{-1}\cdot\vec{\xi}+k_+\epsilon_+^2+k_-\epsilon_-^2+k_m m^2,
\ee
where the levels for $U(1)_m$, $U(1)_x$, and $SU(2)_t$ are given respectively by:
\bea
k_{m}
=
\sum_{a=1}^{r-1}\kappa^{(a)},\qquad
k_-
=
-\frac{1}{2}n\sum_{a,b=1}^{r-1} C^{{A}_{r-1}}_{ab} \kappa^{(a)}\cdot \kappa^{(b)},\qquad
k_+
&=&
-k_--k_m.
\eea

\subsection{Examples}
\label{sec:examples}
\subsubsection{One string on $\mathbb{C}^2/\mathbb{Z}_n$.}
It is straightforward to write down the expression for the elliptic genus of a single M-string on $\mathbb{C}^2/\mathbb{Z}_n$ suspended between the $a$ and $(a+1)$-st NS5 brane, that is, for $\kappa^{(b)}=\delta_{b a}$ . The residues are labeled by a Young diagrams containing exactly $n$ boxes and having the following shape:
\be
\label{eq:yt}
\begin{ytableau}
       ~ & \none & \none & \none\\
       \vdots & \none & \none & \none\\
        & \none & \none & \none \\
       & \cdots  &
\end{ytableau}
\ee
We find that the elliptic genus can be put in the following simple form:
\be
\mathbb{E}^{\boldsymbol{\omega}^{KK}}_{\boldsymbol{\kappa}}(\vec{\xi},\epsilon_+,\epsilon_-,m,\tau) = \sum_{j=1}^{n}\widehat{\chi}^{\mathfrak{su}(n)_1}_{\omega^{KK,(a)}}(\vec{\xi}_+[j])\widehat{\chi}^{\mathfrak{su}(n)_1}_{\omega^{KK,(a+1)}}(\vec{\xi}_-[j])\mathbb{E}(\epsilon_+[j],\epsilon_-[j],m,\tau),
\ee
where
\be
\mathbb{E}(\epsilon_+,\epsilon_-,m,\tau)= -\eta^{-2}\frac{\theta_1(m+\epsilon_+)\theta_1(m-\epsilon_+)}{\theta_1(\epsilon_++\epsilon_-)\theta_1(\epsilon_+-\epsilon_-)}
\ee
is the elliptic genus of one M-string on $\mathbb{C}^2$, and
\be
\xi_{\pm,k}[j] = \xi_k\pm\delta_{k,j}\epsilon_1[j]\pm\delta_{k,j-1}\epsilon_2[j].
\ee
This agrees on the nose with the expression \eqref{eq:ellreso} we obtained in section \ref{sec:comm} by employing from the Nekrasov master formula.\\

The global symmetry on the string includes an unbroken $SU(n)\times SU(n)$, so it should be possible to express the elliptic genus in terms of $\mathfrak{su}(n)\times \mathfrak{su}(n)$ characters with \emph{untwisted} chemical potentials. Indeed, for the case of one M-string on $\mathbb{C}^2/\mathbb{Z}_2$, we find the following alternative expressions for different choices of monodromy (without loss of generality we take $r=2$ here):
\bea
\nonumber
&&
\hskip-.65in
\mathbb{E}^{(0,0)}_{(1)}(\vec{\xi},\epsilon_+,\epsilon_-,m,\tau)
=
-\eta^{-2}\frac{\theta_1(m+\epsilon_+)\theta_1(m-\epsilon_+)}{\theta_1(\epsilon_-)^2}\times\\
\nonumber
&&\bigg[\frac{1}{2}\widehat{\chi}^{\mathfrak{su}(2)_1}_{0}(\vec{\xi})^2\left(\frac{\theta_1(\epsilon_+)^2}{\theta_1(\epsilon_++\epsilon_-)\theta_1(\epsilon_+-\epsilon_-)}-\frac{\theta_2(\epsilon_+)^2}{\theta_2(\epsilon_++\epsilon_-)\theta_2(\epsilon_+-\epsilon_-)}\right)\\
&&+\frac{1}{2}\widehat{\chi}^{\mathfrak{su}(2)_1}_{1}(\vec{\xi})^2\left(\frac{\theta_3(\epsilon_+)^2}{\theta_3(\epsilon_++\epsilon_-)\theta_3(\epsilon_+-\epsilon_-)}-\frac{\theta_4(\epsilon_+)^2}{\theta_4(\epsilon_++\epsilon_-)\theta_4(\epsilon_+-\epsilon_-)}\right)\bigg];
\label{eq:e00}
\eea
\bea
\nonumber
&&
\hskip-.65in
\mathbb{E}^{(1,1)}_{(1)}(\vec{\xi},\epsilon_+,\epsilon_-,m,\tau)
=
-\eta^{-2}\frac{\theta_1(m+\epsilon_+)\theta_1(m-\epsilon_+)}{\theta_1(\epsilon_-)^2}\times\\
\nonumber
&&\bigg[\frac{1}{2}\widehat{\chi}^{\mathfrak{su}(2)_1}_{1}(\vec{\xi})^2\left(\frac{\theta_1(\epsilon_+)^2}{\theta_1(\epsilon_++\epsilon_-)\theta_1(\epsilon_+-\epsilon_-)}-\frac{\theta_2(\epsilon_+)^2}{\theta_2(\epsilon_++\epsilon_-)\theta_2(\epsilon_+-\epsilon_-)}\right)\\
&&+\frac{1}{2}\widehat{\chi}^{\mathfrak{su}(2)_1}_{0}(\vec{\xi})^2\left(\frac{\theta_3(\epsilon_+)^2}{\theta_3(\epsilon_++\epsilon_-)\theta_3(\epsilon_+-\epsilon_-)}-\frac{\theta_4(\epsilon_+)^2}{\theta_4(\epsilon_++\epsilon_-)\theta_4(\epsilon_+-\epsilon_-)}\right)\bigg];
\label{eq:e11}
\eea
and
\bea
\nonumber
&&
\hskip-0.25in
\mathbb{E}^{(0,1)}_{(1)}(\vec{\xi},\epsilon_+,\epsilon_-,m,\tau)=\mathbb{E}^{(1,0)}_{(1)}(\vec{\xi},\epsilon_+,\epsilon_-,m,\tau)
=
-\eta^{-2}\frac{\theta_1(m+\epsilon_+)\theta_1(m-\epsilon_+)}{\theta_1(\epsilon_-)^2}\times\\
\nonumber
&&\frac{1}{2}\widehat{\chi}^{\mathfrak{su}(2)_1}_{0}(\vec{\xi})\widehat{\chi}^{\mathfrak{su}(2)_1}_{1}(\vec{\xi})\times\bigg[\frac{\theta_1(\epsilon_+)^2}{\theta_1(\epsilon_++\epsilon_-)\theta_1(\epsilon_+-\epsilon_-)}+\frac{\theta_2(\epsilon_+)^2}{\theta_2(\epsilon_++\epsilon_-)\theta_2(\epsilon_+-\epsilon_-)}
\\
&&
\nonumber
\hskip1.8in-\frac{\theta_3(\epsilon_+)^2}{\theta_3(\epsilon_++\epsilon_-)\theta_3(\epsilon_+-\epsilon_-)}-\frac{\theta_4(\epsilon_+)^2}{\theta_4(\epsilon_++\epsilon_-)\theta_4(\epsilon_+-\epsilon_-)}\bigg].\\
\eea

\subsubsection{A chain of consecutive M-strings.} We consider the sector in which
\be
\kappa^{(c)} =
\begin{cases}
1 & a\leq c\leq b\\
0 &\text{otherwise}
\end{cases}
,
\ee
corresponding to a bound state of $b-a+1$ M-strings suspended between consecutive NS5 branes. Residues now are labeled by a $(b-a+1)$-tuple of Young diagrams of the shape \eqref{eq:yt}. Also in this case we obtain a suggestive expression for the elliptic genus:
\bea
\nonumber
&&
\hskip-0.7in
\mathbb{E}^{\boldsymbol{\omega}^{KK}}_{\boldsymbol{\kappa}}(\vec{\xi},\epsilon_+,\epsilon_-,m,\tau)
=
\\
\nonumber
&&
\sum_{j_a=1}^{n}
\dots
\sum_{j_b=1}^{n}
\widehat{\chi}^{\mathfrak{su}(n)_1}_{\omega^{KK,(a)}}(\vec{\xi}_+[j_{a}])
\left(\prod_{c=a}^{b-1}\widehat{\chi}^{\mathfrak{su}(n)_1}_{\omega^{KK,(c)}}(\vec{\xi}[j_c,j_{c+1}])\right)
\widehat{\chi}^{\mathfrak{su}(n)_1}_{\omega^{KK,(b)}}(\vec{\xi}_+[j_{b}])
\\
\nonumber
&\times&
\prod_{c=a}^{b-1}\left(\frac{\theta_1(m+\epsilon_-[j_c])\theta_1(m-\epsilon_-[j_c])}{\theta_1(m+\epsilon_+[j_c])\theta_1(m-\epsilon_+[j_c])}\right)^{\delta_{j_c,j_{c+1}}}
\prod_{c=a}^b \mathbb{E}(\epsilon_+[j_c],\epsilon_-[j_c],m,\tau),
\\
\eea
where
\bea
\nonumber
\vec{\xi}[j_c,j_{c+1}]_k = \xi_k&+&\delta_{k,j_{c+1}}(\epsilon_+[j_{c+1}]+\epsilon_-[j_{c+1}])-\delta_{k,j_{c}}(\epsilon_+[j_{c}]+\epsilon_-[j_{c}])
\\
&+&
\delta_{k,j_{c+1}-1}(\epsilon_+[j_{c+1}]-\epsilon_-[j_{c+1}])-\delta_{k,j_{c}-1}(\epsilon_+[j_c]-\epsilon_-[j_c]).
\eea
For a given term in the sum, we can alternatively encode a tuple  $\mathcal{J}=(j_a,j_{a+1},\dots,j_b)$ in terms of a sequence of integer pairs $(\{J_1,n_1\},$ $\{J_2,n_2\},$ $\dots,$ $\{J_{k_\mathcal{J}},n_{k_\mathcal{J}}\})$ such that $J_\ell\neq J_{\ell+1}$, $\sum_{\ell=1}^{k_\mathcal{J}} n_\ell  = b-a+1$, and
\be
\mathcal{J} = (J_1,\dots,J_1,J_2,\dots,J_2,\dots,J_{k_\mathcal{J}},\dots,J_{k_\mathcal{J}})
\ee
where $J_\ell$ is repeated $n_\ell$ times. Then one can write
\bea
\nonumber
\mathbb{E}^{\boldsymbol{\omega}^{KK}}_{\boldsymbol{\kappa}}(\vec{\xi},\epsilon_+,\epsilon_-,m,\tau)
=
\hskip4in
\\
\nonumber
\sum_{j_a=1}^{n}
\dots
\sum_{j_b=1}^{n}
\widehat{\chi}^{\mathfrak{su}(n)_1}_{\omega^{KK,(a)}}(\vec{\xi}_+[j_{a}])
\left(\prod_{c=a}^{b-1}\widehat{\chi}^{\mathfrak{su}(n)_1}_{\omega^{KK,(c)}}(\vec{\xi}[j_c,j_{c+1}])\right)
\widehat{\chi}^{\mathfrak{su}(n)_1}_{\omega^{KK,(b)}}(\vec{\xi}_+[j_{b}])
\\
\nonumber
\times
\prod_{\ell=1}^{k_\mathcal{J}} \mathbb{E}_{n_\ell}(\epsilon_+[j_{J_{\ell}}],\epsilon_-[j_{J_{\ell}}],m,\tau),
\eea
where
\be
\mathbb{E}_{n_\ell}(\epsilon_+,\epsilon_-,m,\tau)
=
\left(\frac{\theta_1(m+\epsilon_-)\theta_1(m-\epsilon_-)}{\theta_1(m+\epsilon_+)\theta_1(m-\epsilon_+)}\right)^{n_\ell-1}
\mathbb{E}(\epsilon_+,\epsilon_-,m,\tau)^{n_\ell}
\ee
is the elliptic genus of a chain of $n_\ell$ consecutive M-strings on $\mathbb{C}^2$. This expression again agrees on the nose with equation \eqref{eq:ellreso}.

\subsubsection{Bound state of two M-strings}
Here we consider the case corresponding to $r=2,\boldsymbol{\kappa} = (2)$, for $n=2,3$.
\paragraph{Two M-strings on $\mathbb{C}^2/\mathbb{Z}_2$}
 The elliptic genus \eqref{eq:ellint} evaluates to a sum over residues which are labeled by all possible Young diagrams of size 4. We find:
\bea
\nonumber
\mathbb{E}^{\boldsymbol{\omega}^{KK}}_{(2)}(\vec{\xi},\epsilon_+,\epsilon_-,m,\tau)
=
\eta^{-2}
\theta_1(m+\epsilon_+)\theta_1(m-\epsilon_+)
\bigg[\hspace{2.4in}
\\
\nonumber
\widehat{\chi}^{\mathfrak{su}(2)_1}_{\omega^{KK,(1)}}
(\xi_{1,+}^{
\resizebox{0.4in}{!}
{\begin{ytableau}
~ & ~ & ~ & ~\\
\end{ytableau}
}})
\widehat{\chi}^{\mathfrak{su}(2)_1}_{\omega^{KK,(2)}}
(\xi_{1,-}^{
\resizebox{0.4in}{!}
{\begin{ytableau}
~ & ~ & ~ & ~\\
\end{ytableau}
}})
\frac{
\theta_1(m-3\epsilon_+-2\epsilon_-)\theta_1(m+3\epsilon_++2\epsilon_-)}{\theta_1(-2\epsilon_-)\theta_1(-2\epsilon_+-4\epsilon_-)\theta_1(2\epsilon_++2\epsilon_-)\theta_1(4\epsilon_++4\epsilon_-)}
\\
\nonumber
+
\widehat{\chi}^{\mathfrak{su}(2)_1}_{\omega^{KK,(1)}}
(\xi_{1,+}^{
\resizebox{0.3in}{!}
{\begin{ytableau}
       ~ \\
       ~ & ~ & ~\\
\end{ytableau}
}})
\widehat{\chi}^{\mathfrak{su}(2)_1}_{\omega^{KK,(2)}}
(\xi_{1,-}^{
\resizebox{0.3in}{!}
{\begin{ytableau}
       ~ \\
       ~ & ~ & ~\\
\end{ytableau}
}})
\frac{
\theta_1(m-3\epsilon_+-2\epsilon_-)\theta_1(m+3\epsilon_++2\epsilon_-)}{\theta_1(-4\epsilon_-)\theta_1(-2\epsilon_-)\theta_1(2\epsilon_++2\epsilon_-)\theta_1(2\epsilon_++4\epsilon_-)}
\\
\nonumber
+
\widehat{\chi}^{\mathfrak{su}(2)_1}_{\omega^{KK,(1)}}
(\xi_{1,+}^{
\resizebox{0.2in}{!}
{\begin{ytableau}
       ~ & ~\\
       ~ & ~\\
\end{ytableau}
}})
\widehat{\chi}^{\mathfrak{su}(2)_1}_{\omega^{KK,(2)}}
(\xi_{1,-}^{
\resizebox{0.2in}{!}
{\begin{ytableau}
       ~ & ~\\
       ~ & ~\\
\end{ytableau}
}})
\frac{\theta_1(m-3\epsilon_+)\theta_1(m+3\epsilon_+)}{\theta_1(-2\epsilon_-)\theta_1(2\epsilon_+-2\epsilon_-)\theta_1(2\epsilon_-)\theta_1(2\epsilon_++2\epsilon_-)}
\\
\nonumber
+
\widehat{\chi}^{\mathfrak{su}(2)_1}_{\omega^{KK,(1)}}
(\xi_{1,+}^{
\resizebox{0.2in}{!}
{\begin{ytableau}
       ~ \\
       ~ \\
       ~ & ~\\
\end{ytableau}
}})
\widehat{\chi}^{\mathfrak{su}(2)_1}_{\omega^{KK,(2)}}
(\xi_{1,-}^{
\resizebox{0.2in}{!}
{\begin{ytableau}
       ~ \\
       ~ \\
       ~ & ~\\
\end{ytableau}
}})
\frac{\theta_1(m-3\epsilon_++2\epsilon_-)\theta_1(m+3\epsilon_+-2\epsilon_-)}{\theta_1(2\epsilon_+-4\epsilon_-)\theta_1(2\epsilon_+-2\epsilon_-)\theta_1(2\epsilon_-)\theta_1(4\epsilon_-)}
\\
\nonumber
+
\widehat{\chi}^{\mathfrak{su}(2)_1}_{\omega^{KK,(1)}}
(\xi_{1,+}^{
\resizebox{0.1in}{!}
{\begin{ytableau}
       ~ \\
       ~ \\
       ~ \\
       ~ \\
\end{ytableau}
}})
\widehat{\chi}^{\mathfrak{su}(2)_1}_{\omega^{KK,(2)}}
(\xi_{1,-}^{
\resizebox{0.1in}{!}
{\begin{ytableau}
       ~ \\
       ~ \\
       ~ \\
       ~ \\
\end{ytableau}
}})
\frac{\theta_1(m-3\epsilon_++2\epsilon_-)\theta_1(m+3\epsilon_+-2\epsilon_-)}{\theta_1(4\epsilon_+-4\epsilon_-)\theta_1(2\epsilon_+-2\epsilon_-)\theta_1(2\epsilon_-)\theta_1(-2\epsilon_++4\epsilon_-)}
\bigg]
\\
\label{eq:2strres}
\eea
where
\be
\nonumber
\xi_{1,\pm}^{
\resizebox{0.4in}{!}
{\begin{ytableau}
       ~ & ~ & ~ & ~\\
\end{ytableau}
}}
= \xi_1\pm 4(\epsilon_++\epsilon_-)
,\qquad
\xi_{1,\pm}^{
\resizebox{0.3in}{!}
{\begin{ytableau}
       ~ \\
       ~ & ~ & ~\\
\end{ytableau}
}}
= \xi_1\mp 4\epsilon_-
,\qquad
\xi_{1,\pm}^{
\resizebox{0.2in}{!}
{\begin{ytableau}
       ~ & ~\\
       ~ & ~\\
\end{ytableau}
}}
= \xi_1,
\ee
\be
\xi_{1,\pm}^{
\resizebox{0.2in}{!}
{\begin{ytableau}
       ~ \\
       ~ \\
       ~ & ~\\
\end{ytableau}
}}
= \xi_1\pm 4\epsilon_-
,\qquad
\xi_{1,\pm}^{
\resizebox{0.1in}{!}
{\begin{ytableau}
       ~ \\
       ~ \\
       ~ \\
       ~ \\
\end{ytableau}
}}
= \xi_1\pm 4(\epsilon_+-\epsilon_-).
\ee
On the other hand, equation \eqref{eq:ellreso} gives rise to the following expression for the elliptic genus on $\widetilde{\mathbb{C}^2/\mathbb{Z}_2}$:
\bea
\nonumber
\mathbb{E}^{\boldsymbol{\omega}^{KK}}_{(2)}(\vec{\xi},\epsilon_+,\epsilon_-,m,\tau)
=
\frac{\theta_1(m+\epsilon_+)\theta_1(m-\epsilon_+)}{\eta^{2}\,\theta_1(2\epsilon_-)}
\bigg[\frac{\widehat{\chi}^{\mathfrak{su}(2)_1}_{\omega^{KK,(1)}}
(\xi_{1}+4\epsilon_1)
\widehat{\chi}^{\mathfrak{su}(2)_1}_{\omega^{KK,(2)}}
(\xi_{1}-4\epsilon_1)}{\theta_1(2\epsilon_1)}
\\
\nonumber
\left(
\frac{
\theta_1(m+\epsilon_+-2\epsilon_-)
\theta_1(m-\epsilon_++2\epsilon_-)}
{\theta_1(2\epsilon_++4\epsilon_-)
\theta_1(4\epsilon_-)}+
\frac{
\theta_1(m-3\epsilon_+-2\epsilon_-)
\theta_1(m+3\epsilon_++2\epsilon_-)}
{\theta_1(2\epsilon_++4\epsilon_-)
\theta_1(4\epsilon_++4\epsilon_-)}
\right)
\\
\nonumber
+\frac{\widehat{\chi}^{\mathfrak{su}(2)_1}_{\omega^{KK,(1)}}
(\xi_{1}+4\epsilon_2)
\widehat{\chi}^{\mathfrak{su}(2)_1}_{\omega^{KK,(2)}}
(\xi_{1}-4\epsilon_2)}{\theta_1(2\epsilon_2)}
\bigg(
\frac{
\theta_1(m-\epsilon_+-2\epsilon_-)
\theta_1(m+\epsilon_++2\epsilon_-)}
{\theta_1(2\epsilon_+-4\epsilon_-)
\theta_1(4\epsilon_-)}+
\\
\nonumber
\frac{
\theta_1(m+3\epsilon_+-2\epsilon_-)
\theta_1(m-3\epsilon_++2\epsilon_-)}
{\theta_1(4\epsilon_+-4\epsilon_-)
\theta_1(-2\epsilon_++4\epsilon_-)}
\bigg)
+
\frac{\widehat{\chi}^{\mathfrak{su}(2)_1}_{\omega^{KK,(1)}}
(\xi_{1}+4\epsilon_+)
\widehat{\chi}^{\mathfrak{su}(2)_1}_{\omega^{KK,(2)}}
(\xi_{1}-4\epsilon_+)}{\theta_1(2\epsilon_2)}
\\
\nonumber
\times
\frac{\theta_1(m+\epsilon_+)\theta_1(m-\epsilon_+)}{\theta_1(-2\epsilon_-)\theta_1(2\epsilon_1)\theta_1(2\epsilon_2)}\bigg].\hspace{3.5in}\\
\label{eq:2strfac}
\eea
Although equations \eqref{eq:2strres} and \eqref{eq:2strfac} are not obviously equal to each other, in fact one can verify by expanding in powers of $e^{2\pi i \tau}$ that they coincide.

\paragraph{Two M-strings on $\mathbb{C}^2/\mathbb{Z}_3$}
 In this case the residues which are labeled by all possible Young diagrams of size 6 except for the following two diagrams:
\be
\nonumber
\resizebox{0.4in}{!}{
\begin{ytableau}
       ~ \\
       ~ \\
       ~ & \\
       ~ & \\
\end{ytableau}}
\qquad\qquad
\resizebox{0.8in}{!}{
\begin{ytableau}
\none\\
       ~ & ~ \\
       ~ & ~ & ~ & ~ \\
\end{ytableau}}
\ee
Summing over residues, we obtain the following expression for the elliptic genus on $\mathbb{C}^2/\mathbb{Z}_3$:
\bea
\nonumber
\mathbb{E}^{\boldsymbol{\omega}^{KK}}_{(2)}(\vec{\xi},\epsilon_+,\epsilon_-,m,\tau)
=
\eta^{-2}
\theta_1(m+\epsilon_+)\theta_1(m-\epsilon_+)
\bigg[\hspace{2.3in}
\\
\nonumber
\resizebox{\textwidth}{!}{
$
\widehat{\chi}^{\mathfrak{su}(2)_1}_{\omega^{KK,(1)}}
(\vec{\xi}_{+}^{
\resizebox{0.6in}{!}
{\begin{ytableau}
~ & ~ & ~ & ~ & ~ & ~\\
\end{ytableau}
}})
\widehat{\chi}^{\mathfrak{su}(2)_1}_{\omega^{KK,(2)}}
(\vec{\xi}_{-}^{
\resizebox{0.6in}{!}
{\begin{ytableau}
~ & ~ & ~ & ~ & ~ & ~\\
\end{ytableau}
}})
\frac{
\theta_1(m-4\epsilon_+-3\epsilon_-)\theta_1(m+4\epsilon_++3\epsilon_-)}{\theta_1(-4\epsilon_+-6\epsilon_-)\theta_1(-\epsilon_+-3\epsilon_-)\theta_1(3\epsilon_++3\epsilon_-)\theta_1(6\epsilon_++6\epsilon_-)}
$
}
\\
\nonumber
\resizebox{\textwidth}{!}{
$
+
\widehat{\chi}^{\mathfrak{su}(2)_1}_{\omega^{KK,(1)}}
(\vec{\xi}_{+}^{
\resizebox{0.5in}{!}
{\begin{ytableau}
       ~ \\
       ~ & ~ & ~ & ~ & ~\\
\end{ytableau}
}})
\widehat{\chi}^{\mathfrak{su}(2)_1}_{\omega^{KK,(2)}}
(\vec{\xi}_{-}^{
\resizebox{0.5in}{!}
{\begin{ytableau}
       ~ \\
       ~ & ~ & ~ & ~ & ~\\
\end{ytableau}
}})
\frac{
\theta_1(m-4\epsilon_+-3\epsilon_-)\theta_1(m+4\epsilon_++3\epsilon_-)}{\theta_1(-2\epsilon_+-6\epsilon_-)\theta_1(-\epsilon_+-3\epsilon_-)\theta_1(3\epsilon_++3\epsilon_-)\theta_1(4\epsilon_++6\epsilon_-)}
$
}
\\
\nonumber
\resizebox{0.9\textwidth}{!}{
$
+
\widehat{\chi}^{\mathfrak{su}(2)_1}_{\omega^{KK,(1)}}
(\vec{\xi}_{+}^{
\resizebox{0.4in}{!}
{\begin{ytableau}
       ~ \\
       ~ \\
       ~ & ~ & ~ & ~\\
\end{ytableau}
}})
\widehat{\chi}^{\mathfrak{su}(2)_1}_{\omega^{KK,(2)}}
(\vec{\xi}_{-}^{
\resizebox{0.4in}{!}
{\begin{ytableau}
       ~ \\
       ~ \\
       ~ & ~ & ~ & ~\\
\end{ytableau}
}})
\frac{\theta_1(m-4\epsilon_+-3\epsilon_-)\theta_1(m+4\epsilon_++3\epsilon_-)}{\theta_1(-6\epsilon_-)\theta_1(-\epsilon_+-3\epsilon_-)\theta_1(3\epsilon_++3\epsilon_-)\theta_1(2\epsilon_++6\epsilon_-)}
$
}
\\
\nonumber
\resizebox{0.85\textwidth}{!}{
$
+
\widehat{\chi}^{\mathfrak{su}(2)_1}_{\omega^{KK,(1)}}
(\vec{\xi}_{+}^{
\resizebox{0.3in}{!}
{\begin{ytableau}
       ~ & ~ & ~\\
       ~ & ~ & ~\\
\end{ytableau}
}})
\widehat{\chi}^{\mathfrak{su}(2)_1}_{\omega^{KK,(2)}}
(\vec{\xi}_{-}^{
\resizebox{0.3in}{!}
{\begin{ytableau}
       ~ & ~ & ~\\
       ~ & ~ & ~\\
\end{ytableau}
}})
\frac{\theta_1(m-3\epsilon_+)\theta_1(m+3\epsilon_+)}{\theta_1(-\epsilon_+-3\epsilon_-)\theta_1(\epsilon_+-3\epsilon_-)\theta_1(\epsilon_++3\epsilon_-)\theta_1(3\epsilon_++3\epsilon_-)}
$
}
\\
\nonumber
\resizebox{0.6\textwidth}{!}{
$
+
\widehat{\chi}^{\mathfrak{su}(2)_1}_{\omega^{KK,(1)}}
(\vec{\xi}_{+}^{
\resizebox{0.3in}{!}
{\begin{ytableau}
       ~ \\
       ~ & ~\\
       ~ & ~ & ~\\
\end{ytableau}
}})
\widehat{\chi}^{\mathfrak{su}(2)_1}_{\omega^{KK,(2)}}
(\vec{\xi}_{-}^{
\resizebox{0.3in}{!}
{\begin{ytableau}
       ~ \\
       ~ & ~\\
       ~ & ~ & ~\\
\end{ytableau}
}})
\frac{\theta_1(m-3\epsilon_+)\theta_1(m+3\epsilon_+)}{\theta_1(\epsilon_+-3\epsilon_-)^2\theta_1(\epsilon_++3\epsilon_-)^2}
$
}
\\
\nonumber
\resizebox{0.8\textwidth}{!}{
$
+
\widehat{\chi}^{\mathfrak{su}(2)_1}_{\omega^{KK,(1)}}
(\vec{\xi}_{+}^{
\resizebox{0.3in}{!}
{\begin{ytableau}
       ~ \\
       ~ \\
       ~ \\
       ~ & ~ & ~\\
\end{ytableau}
}})
\widehat{\chi}^{\mathfrak{su}(2)_1}_{\omega^{KK,(2)}}
(\vec{\xi}_{-}^{
\resizebox{0.3in}{!}
{\begin{ytableau}
       ~ \\
       ~ \\
       ~ \\
       ~ & ~ & ~\\
\end{ytableau}
}})
\frac{\theta_1(m+4\epsilon_+-3\epsilon_-)\theta_1(m-4\epsilon_++3\epsilon_-)}{\theta_1(2\epsilon_+-6\epsilon_-)\theta_1(3\epsilon_+-3\epsilon_-)\theta_1(-\epsilon_++3\epsilon_-)\theta_1(6\epsilon_-)}
$
}
\\
\nonumber
\resizebox{0.8\textwidth}{!}{
$
+
\widehat{\chi}^{\mathfrak{su}(2)_1}_{\omega^{KK,(1)}}
(\vec{\xi}_{+}^{
\resizebox{0.2in}{!}
{\begin{ytableau}
       ~ & ~ \\
       ~ & ~ \\
       ~ & ~ \\
\end{ytableau}
}})
\widehat{\chi}^{\mathfrak{su}(2)_1}_{\omega^{KK,(2)}}
(\vec{\xi}_{-}^{
\resizebox{0.2in}{!}
{\begin{ytableau}
       ~ & ~ \\
       ~ & ~ \\
       ~ & ~ \\
\end{ytableau}
}})
\frac{\theta_1(m-3\epsilon_+)\theta_1(m+3\epsilon_+)}{\theta_1(\epsilon_+-3\epsilon_-)\theta_1(3\epsilon_+-3\epsilon_-)\theta_1(-\epsilon_++3\epsilon_-)\theta_1(\epsilon_++3\epsilon_-)}
$
}
\\
\nonumber
\resizebox{0.8\textwidth}{!}{
$
+
\widehat{\chi}^{\mathfrak{su}(2)_1}_{\omega^{KK,(1)}}
(\vec{\xi}_{+}^{
\resizebox{0.2in}{!}
{\begin{ytableau}
       ~ \\
       ~ \\
       ~ \\
       ~ \\
       ~ & ~\\
\end{ytableau}
}})
\widehat{\chi}^{\mathfrak{su}(2)_1}_{\omega^{KK,(2)}}
(\vec{\xi}_{-}^{
\resizebox{0.2in}{!}
{\begin{ytableau}
       ~ \\
       ~ \\
       ~ \\
       ~ \\
       ~ & ~\\
\end{ytableau}
}})
\frac{\theta_1(m+4\epsilon_+-3\epsilon_-)\theta_1(m-4\epsilon_++3\epsilon_-)}{\theta_1(4\epsilon_+-6\epsilon_-)\theta_1(3\epsilon_+-3\epsilon_-)\theta_1(-\epsilon_++3\epsilon_-)\theta_1(-2\epsilon_++6\epsilon_-)}
$
}
\\
\nonumber
\resizebox{0.8\textwidth}{!}{
$
+
\widehat{\chi}^{\mathfrak{su}(2)_1}_{\omega^{KK,(1)}}
(\vec{\xi}_{+}^{
\resizebox{0.1in}{!}
{\begin{ytableau}
       ~ \\
       ~ \\
       ~ \\
       ~ \\
       ~ \\
       ~ \\
\end{ytableau}
}})
\widehat{\chi}^{\mathfrak{su}(2)_1}_{\omega^{KK,(2)}}
(\vec{\xi}_{-}^{
\resizebox{0.1in}{!}
{\begin{ytableau}
       ~ \\
       ~ \\
       ~ \\
       ~ \\
       ~ \\
       ~ \\
\end{ytableau}
}})
\frac{\theta_1(m+4\epsilon_+-3\epsilon_-)\theta_1(m-4\epsilon_++3\epsilon_-)}{\theta_1(6\epsilon_+-6\epsilon_-)\theta_1(3\epsilon_+-3\epsilon_-)\theta_1(-\epsilon_++3\epsilon_-)\theta_1(-4\epsilon_++6\epsilon_-)}
$
}
\bigg]
\\
\label{eq:elln3r}
\eea
where
\bea
\nonumber
\xi_{1,\pm}^{
\resizebox{0.6in}{!}
{\begin{ytableau}
~ & ~ & ~ & ~ & ~ & ~\\
\end{ytableau}
}}
=
\xi_1\pm 6(\epsilon_++\epsilon_-)
,
\qquad
\xi_{1,\pm}^{
\resizebox{0.5in}{!}
{\begin{ytableau}
       ~ \\
       ~ & ~ & ~ & ~ & ~\\
\end{ytableau}
}}
=
\xi_1\mp (2\epsilon_++6\epsilon_-)
,
\qquad
\xi_{1,\pm}^{
\resizebox{0.4in}{!}
{\begin{ytableau}
       ~ \\
       ~ \\
       ~ & ~ & ~ & ~\\
\end{ytableau}
}}
=
\xi_1,
\\
\nonumber
\xi_{2,\pm}^{
\resizebox{0.6in}{!}
{\begin{ytableau}
~ & ~ & ~ & ~ & ~ & ~\\
\end{ytableau}
}}
=
\xi_2
,
\qquad
\xi_{2,\pm}^{
\resizebox{0.5in}{!}
{\begin{ytableau}
       ~ \\
       ~ & ~ & ~ & ~ & ~\\
\end{ytableau}
}}
=
\xi_2\pm (4\epsilon_++6\epsilon_-)
,
\qquad
\xi_{2,\pm}^{
\resizebox{0.4in}{!}
{\begin{ytableau}
       ~ \\
       ~ \\
       ~ & ~ & ~ & ~\\
\end{ytableau}
}}
=
\xi_2\mp 6\epsilon_-,
\\
\nonumber
\xi_{1,\pm}^{
\resizebox{0.3in}{!}
{\begin{ytableau}
       ~ & ~ & ~\\
       ~ & ~ & ~\\
\end{ytableau}
}}
=
\xi_1
,
\qquad
\xi_{1,\pm}^{
\resizebox{0.3in}{!}
{\begin{ytableau}
       ~ \\
       ~ & ~\\
       ~ & ~ & ~\\
\end{ytableau}
}}
=
\xi_1\pm (\epsilon_++3\epsilon_-)
,
\qquad
\xi_{1,\pm}^{
\resizebox{0.3in}{!}
{\begin{ytableau}
       ~ \\
       ~ \\
       ~ \\
       ~ & ~ & ~\\
\end{ytableau}
}}
=
\xi_1\pm 6\epsilon_-,
\\
\nonumber
\xi_{2,\pm}^{
\resizebox{0.3in}{!}
{\begin{ytableau}
       ~ & ~ & ~\\
       ~ & ~ & ~\\
\end{ytableau}
}}
=
\xi_2\pm 3(\epsilon_++\epsilon_-)
,
\qquad
\xi_{2,\pm}^{
\resizebox{0.3in}{!}
{\begin{ytableau}
       ~ \\
       ~ & ~\\
       ~ & ~ & ~\\
\end{ytableau}
}}
=
\xi_2\pm (\epsilon_+-3\epsilon_-)
,
\qquad
\xi_{2,\pm}^{
\resizebox{0.3in}{!}
{\begin{ytableau}
       ~ \\
       ~ \\
       ~ \\
       ~ & ~ & ~\\
\end{ytableau}
}}
=
\xi_2,
\\
\nonumber
\xi_{1,\pm}^{
\resizebox{0.2in}{!}
{\begin{ytableau}
       ~ & ~ \\
       ~ & ~ \\
       ~ & ~ \\
\end{ytableau}
}}
=
\xi_1\pm 3(\epsilon_+-\epsilon_-)
,
\qquad
\xi_{1,\pm}^{
\resizebox{0.2in}{!}
{\begin{ytableau}
       ~ \\
       ~ \\
       ~ \\
       ~ \\
       ~ & ~\\
\end{ytableau}
}}
=
\xi_1\pm (4\epsilon_+-6\epsilon_-)
,
\qquad
\xi_{1,\pm}^{
\resizebox{0.1in}{!}
{\begin{ytableau}
       ~ \\
       ~ \\
       ~ \\
       ~ \\
       ~ \\
       ~ \\
\end{ytableau}
}}
=
\xi_1,
\\
\nonumber
\xi_{2,\pm}^{
\resizebox{0.2in}{!}
{\begin{ytableau}
       ~ & ~ \\
       ~ & ~ \\
       ~ & ~ \\
\end{ytableau}
}}
=
\xi_2
,
\qquad
\xi_{2,\pm}^{
\resizebox{0.2in}{!}
{\begin{ytableau}
       ~ \\
       ~ \\
       ~ \\
       ~ \\
       ~ & ~\\
\end{ytableau}
}}
=
\xi_2\mp (2\epsilon_+-6\epsilon_-)
,
\qquad
\xi_{2,\pm}^{
\resizebox{0.1in}{!}
{\begin{ytableau}
       ~ \\
       ~ \\
       ~ \\
       ~ \\
       ~ \\
       ~ \\
\end{ytableau}
}}
=
\xi_2\pm 6(\epsilon_+-\epsilon_-).
\eea
Again formula \eqref{eq:ellreso} gives rise to a seemingly different combination of theta functions and $\mathfrak{su}(n)_1$ characters. By by explicit comparison of the series expansions, we find in fact that it coincides with equation \eqref{eq:elln3r}.
\subsubsection{Bound state of three M-strings on $\mathbb{C}^2/\mathbb{Z}_2$}
We now turn to the case $r=n=2$, $\boldsymbol{\kappa}=(3)$. The residues of equation \eqref{eq:ellint} correspond to all possible Young diagrams of size six, with the exception of:
\be
\nonumber
\resizebox{0.4in}{!}{
\begin{ytableau}
       ~ \\
       ~ & ~\\
       ~ & ~ & ~\\
\end{ytableau}}.
\ee
The elliptic genus is given by:
\bea
\nonumber
&&\mathbb{E}^{\boldsymbol{\omega}^{KK}}_{(3)}(\vec{\xi},\epsilon_+,\epsilon_-,m,\tau)
=
-
\eta^{-2}
\frac{\theta_1(m+\epsilon_+)\theta_1(m-\epsilon_+)}{\theta_1(2\epsilon_-)}
\bigg[\hspace{3in}
\\
\nonumber
&&\hspace{-.1in}
\resizebox{\textwidth}{!}{
$
\widehat{\chi}^{\mathfrak{su}(2)_1}_{\omega^{KK,(1)}}
(\xi_{1,+}^{
\resizebox{0.6in}{!}
{\begin{ytableau}
~ & ~ & ~ & ~ & ~ & ~\\
\end{ytableau}
}})
\widehat{\chi}^{\mathfrak{su}(2)_1}_{\omega^{KK,(2)}}
(\xi_{1,-}^{
\resizebox{0.6in}{!}
{\begin{ytableau}
~ & ~ & ~ & ~ & ~ & ~\\
\end{ytableau}
}})
\frac{
\theta_1(m-5\epsilon_+-4\epsilon_-)
\theta_1(m-3\epsilon_+-2\epsilon_-)
\theta_1(m+3\epsilon_++2\epsilon_-)
\theta_1(m+5\epsilon_++4\epsilon_-)}
{
\theta_1(-4\epsilon_+-6\epsilon_-)
\theta_1(2\epsilon_++4\epsilon_-)
\theta_1(2\epsilon_++2\epsilon_-)
\theta_1(4\epsilon_++4\epsilon_-)
\theta_1(6\epsilon_++6\epsilon_-)
}
$
}
\\
\nonumber
&&
\hspace{-.1in}
\resizebox{\textwidth}{!}{
$
+
\widehat{\chi}^{\mathfrak{su}(2)_1}_{\omega^{KK,(1)}}
(\xi_{1,+}^{
\resizebox{0.5in}{!}
{\begin{ytableau}
       ~ \\
       ~ & ~ & ~ & ~ & ~\\
\end{ytableau}
}})
\widehat{\chi}^{\mathfrak{su}(2)_1}_{\omega^{KK,(2)}}
(\xi_{1,-}^{
\resizebox{0.5in}{!}
{\begin{ytableau}
       ~ \\
       ~ & ~ & ~ & ~ & ~\\
\end{ytableau}
}})
\frac{
\theta_1(m-5\epsilon_+-4\epsilon_-)
\theta_1(m-3\epsilon_+-2\epsilon_-)
\theta_1(m+3\epsilon_++2\epsilon_-)
\theta_1(m+5\epsilon_++4\epsilon_-)}
{
\theta_1(-2\epsilon_+-6\epsilon_-)
\theta_1(2\epsilon_++4\epsilon_-)
\theta_1(2\epsilon_++2\epsilon_-)
\theta_1(4\epsilon_++4\epsilon_-)
\theta_1(4\epsilon_++6\epsilon_-)}
$
}
\\
\nonumber
&&\hspace{-.1in}
\resizebox{\textwidth}{!}{
$
+
\widehat{\chi}^{\mathfrak{su}(2)_1}_{\omega^{KK,(1)}}
(\xi_{1,+}^{
\resizebox{0.4in}{!}
{\begin{ytableau}
       ~ & ~ \\
       ~ & ~ & ~ & ~\\
\end{ytableau}
}})
\widehat{\chi}^{\mathfrak{su}(2)_1}_{\omega^{KK,(2)}}
(\xi_{1,-}^{
\resizebox{0.4in}{!}
{\begin{ytableau}
       ~ & ~ \\
       ~ & ~ & ~ & ~\\
\end{ytableau}
}})
\frac{
\theta_1(m-3\epsilon_+-2\epsilon_-)
\theta_1(m-3\epsilon_+)
\theta_1(m+3\epsilon_+)
\theta_1(m+3\epsilon_++2\epsilon_-)}
{
\theta_1(-4\epsilon_-)
\theta_1(2\epsilon_-)
\theta_1(2\epsilon_++2\epsilon_-)^2
\theta_1(2\epsilon_++4\epsilon_-)
}
\hspace{.4in}
$
}
\\
\nonumber
&&\hspace{-.1in}
\resizebox{\textwidth}{!}{
$
+
\widehat{\chi}^{\mathfrak{su}(2)_1}_{\omega^{KK,(1)}}
(\xi_{1,+}^{
\resizebox{0.4in}{!}
{\begin{ytableau}
       ~ \\
       ~ \\
       ~ & ~ & ~ & ~\\
\end{ytableau}
}})
\widehat{\chi}^{\mathfrak{su}(2)_1}_{\omega^{KK,(2)}}
(\xi_{1,-}^{
\resizebox{0.4in}{!}
{\begin{ytableau}
       ~ \\
       ~ \\
       ~ & ~ & ~ & ~\\
\end{ytableau}
}})
\frac{
\theta_1(m-3\epsilon_+-2\epsilon_-)
\theta_1(m+3\epsilon_+-2\epsilon_-)
\theta_1(m-3\epsilon_++2\epsilon_-)
\theta_1(m+3\epsilon_++2\epsilon_-)}
{
\theta_1(-6\epsilon_-)
\theta_1(-2\epsilon_-)
\theta_1(2\epsilon_+-2\epsilon_-)
\theta_1(2\epsilon_++2\epsilon_-)
\theta_1(2\epsilon_++6\epsilon_-)}
$
}
\\
\nonumber
&&\hspace{-.1in}
\resizebox{.9\textwidth}{!}{$
+
\widehat{\chi}^{\mathfrak{su}(2)_1}_{\omega^{KK,(1)}}
(\xi_{1,+}^{
\resizebox{0.3in}{!}
{\begin{ytableau}
       ~ & ~ & ~\\
       ~ & ~ & ~\\
\end{ytableau}
}})
\widehat{\chi}^{\mathfrak{su}(2)_1}_{\omega^{KK,(2)}}
(\xi_{1,-}^{
\resizebox{0.3in}{!}
{\begin{ytableau}
       ~ & ~ & ~\\
       ~ & ~ & ~\\
\end{ytableau}
}})
\frac{
\theta_1(m-3\epsilon_+-2\epsilon_-)
\theta_1(m-3\epsilon_+)
\theta_1(m+3\epsilon_+)
\theta_1(m+3\epsilon_++2\epsilon_-)}
{
\theta_1(-4\epsilon_-)
\theta_1(-2\epsilon_-)
\theta_1(2\epsilon_+-2\epsilon_-)
\theta_1(2\epsilon_++2\epsilon_-)
\theta_1(2\epsilon_++4\epsilon_-)}
$
}
\\
\nonumber
&&\hspace{-.1in}
\resizebox{\textwidth}{!}{
$
+
\widehat{\chi}^{\mathfrak{su}(2)_1}_{\omega^{KK,(1)}}
(\xi_{1,+}^{
\resizebox{0.3in}{!}
{\begin{ytableau}
       ~ \\
       ~ \\
       ~ \\
       ~ & ~ & ~\\
\end{ytableau}
}})
\widehat{\chi}^{\mathfrak{su}(2)_1}_{\omega^{KK,(2)}}
(\xi_{1,-}^{
\resizebox{0.3in}{!}
{\begin{ytableau}
       ~ \\
       ~ \\
       ~ \\
       ~ & ~ & ~\\
\end{ytableau}
}})
\frac{
\theta_1(m-3\epsilon_+-2\epsilon_-)
\theta_1(m+3\epsilon_+-2\epsilon_-)
\theta_1(m-3\epsilon_++2\epsilon_-)
\theta_1(m+3\epsilon_++2\epsilon_-)}
{
\theta_1(2\epsilon_+-6\epsilon_-)
\theta_1(-2\epsilon_-)
\theta_1(2\epsilon_+-2\epsilon_-)
\theta_1(2\epsilon_++2\epsilon_-)
\theta_1(6\epsilon_-)}
$
}
\\
\nonumber
&&\hspace{-.1in}
\resizebox{.9\textwidth}{!}{
$
+
\widehat{\chi}^{\mathfrak{su}(2)_1}_{\omega^{KK,(1)}}
(\xi_{1,+}^{
\resizebox{0.2in}{!}
{\begin{ytableau}
       ~ & ~ \\
       ~ & ~ \\
       ~ & ~ \\
\end{ytableau}
}})
\widehat{\chi}^{\mathfrak{su}(2)_1}_{\omega^{KK,(2)}}
(\xi_{1,-}^{
\resizebox{0.2in}{!}
{\begin{ytableau}
       ~ & ~ \\
       ~ & ~ \\
       ~ & ~ \\
\end{ytableau}
}})
\frac{
\theta_1(m-3\epsilon_++2\epsilon_-)
\theta_1(m-3\epsilon_+)
\theta_1(m+3\epsilon_+)
\theta_1(m+3\epsilon_+-2\epsilon_-)}
{
\theta_1(2\epsilon_+-4\epsilon_-)
\theta_1(-2\epsilon_-)
\theta_1(2\epsilon_+-2\epsilon_-)
\theta_1(2\epsilon_++2\epsilon_-)
\theta_1(4\epsilon_-)}
$
}
\\
\nonumber
&&\hspace{-.1in}
\resizebox{\textwidth}{!}{
$
+
\widehat{\chi}^{\mathfrak{su}(2)_1}_{\omega^{KK,(1)}}
(\xi_{1,+}^{
\resizebox{0.2in}{!}
{\begin{ytableau}
       ~ \\
       ~ \\
       ~ & ~ \\
       ~ & ~ \\
\end{ytableau}
}})
\widehat{\chi}^{\mathfrak{su}(2)_1}_{\omega^{KK,(2)}}
(\xi_{1,-}^{
\resizebox{0.2in}{!}
{\begin{ytableau}
       ~ \\
       ~ \\
       ~ & ~ \\
       ~ & ~ \\
\end{ytableau}
}})
\frac{
\theta_1(m-3\epsilon_++2\epsilon_-)
\theta_1(m-3\epsilon_+)
\theta_1(m+3\epsilon_+)
\theta_1(m+3\epsilon_+-2\epsilon_-)}
{
\theta_1(2\epsilon_+-4\epsilon_-)
\theta_1(2\epsilon_+-2\epsilon_-)^2
\theta_1(2\epsilon_-)
\theta_1(4\epsilon_-)}
\hspace{.8in}
$
}
\\
\nonumber
&&\hspace{-.1in}
\resizebox{\textwidth}{!}{
$
+
\widehat{\chi}^{\mathfrak{su}(2)_1}_{\omega^{KK,(1)}}
(\xi_{1,+}^{
\resizebox{0.2in}{!}
{\begin{ytableau}
       ~ \\
       ~ \\
       ~ \\
       ~ \\
       ~ & ~\\
\end{ytableau}
}})
\widehat{\chi}^{\mathfrak{su}(2)_1}_{\omega^{KK,(2)}}
(\xi_{1,-}^{
\resizebox{0.2in}{!}
{\begin{ytableau}
       ~ \\
       ~ \\
       ~ \\
       ~ \\
       ~ & ~\\
\end{ytableau}
}})
\frac{
\theta_1(m-5\epsilon_++4\epsilon_-)
\theta_1(m-3\epsilon_++2\epsilon_-)
\theta_1(m+3\epsilon_+-2\epsilon_-)
\theta_1(m+5\epsilon_+-4\epsilon_-)}
{
\theta_1(4\epsilon_+-6\epsilon_-)
\theta_1(4\epsilon_+-4\epsilon_-)
\theta_1(2\epsilon_+-2\epsilon_-)
\theta_1(-2\epsilon_++4\epsilon_-)
\theta_1(-2\epsilon_++6\epsilon_-)}
$
}
\\
\nonumber
&&\hspace{-.1in}
\resizebox{\textwidth}{!}{
$
+
\widehat{\chi}^{\mathfrak{su}(2)_1}_{\omega^{KK,(1)}}
(\xi_{1,+}^{
\resizebox{0.1in}{!}
{\begin{ytableau}
       ~ \\
       ~ \\
       ~ \\
       ~ \\
       ~ \\
       ~ \\
\end{ytableau}
}})
\widehat{\chi}^{\mathfrak{su}(2)_1}_{\omega^{KK,(2)}}
(\xi_{1,-}^{
\resizebox{0.1in}{!}
{\begin{ytableau}
       ~ \\
       ~ \\
       ~ \\
       ~ \\
       ~ \\
       ~ \\
\end{ytableau}
}})
\frac{
\theta_1(m-5\epsilon_++4\epsilon_-)
\theta_1(m-3\epsilon_++2\epsilon_-)
\theta_1(m+3\epsilon_+-2\epsilon_-)
\theta_1(m+5\epsilon_+-4\epsilon_-)}
{
\theta_1(6\epsilon_+-6\epsilon_-)
\theta_1(4\epsilon_+-4\epsilon_-)
\theta_1(2\epsilon_+-2\epsilon_-)
\theta_1(-2\epsilon_++4\epsilon_-)
\theta_1(-4\epsilon_++6\epsilon_-)}
\bigg]
$
}
\eea
\be
\phantom{.}
\label{eq:3str}
\ee
where
\bea
\nonumber
\resizebox{\textwidth}{!}{
$
\xi_{1,\pm}^{
\resizebox{0.6in}{!}
{\begin{ytableau}
~ & ~ & ~ & ~ & ~ & ~\\
\end{ytableau}
}}
=
\xi_1\pm 6(\epsilon_++\epsilon_-)
,
\quad
\xi_{1,\pm}^{
\resizebox{0.5in}{!}
{\begin{ytableau}
       ~ \\
       ~ & ~ & ~ & ~ & ~\\
\end{ytableau}
}}
=
\xi_1\mp (2\epsilon_++6\epsilon_-)
,
\quad
\xi_{1,\pm}^{
\resizebox{0.4in}{!}
{\begin{ytableau}
       ~ & ~ \\
       ~ & ~ & ~ & ~\\
\end{ytableau}
}}
=
\xi_1\pm2(\epsilon_++\epsilon_-),
$
}
\\
\nonumber
\resizebox{\textwidth}{!}{
$
\xi_{1,\pm}^{
\resizebox{0.4in}{!}
{\begin{ytableau}
       ~ \\
       ~ \\
       ~ & ~ & ~ & ~\\
\end{ytableau}
}}
=
\xi_1\pm(2\epsilon_++6\epsilon_-)
,
\qquad
\xi_{1,\pm}^{
\resizebox{0.3in}{!}
{\begin{ytableau}
       ~ & ~ & ~\\
       ~ & ~ & ~\\
\end{ytableau}
}}
=
\xi_1\pm2(\epsilon_+-\epsilon_-)
,
\qquad
\xi_{1,\pm}^{
\resizebox{0.3in}{!}
{\begin{ytableau}
       ~ \\
       ~ \\
       ~ \\
       ~ & ~ & ~\\
\end{ytableau}
}}
=
\xi_1\pm (2\epsilon_+-6\epsilon_-),
$
}
\\
\nonumber
\resizebox{\textwidth}{!}{
$
\xi_{1,\pm}^{
\resizebox{0.2in}{!}
{\begin{ytableau}
       ~ & ~ \\
       ~ & ~ \\
       ~ & ~ \\
\end{ytableau}
}}
=
\xi_1\pm 2(\epsilon_++\epsilon_-)
,
\quad
\xi_{1,\pm}^{
\resizebox{0.2in}{!}
{\begin{ytableau}
       ~ \\
       ~ \\
       ~ & ~\\
       ~ & ~\\
\end{ytableau}
}}
=
\xi_1\pm 2(\epsilon_+-\epsilon_-)
,
\quad
\xi_{1,\pm}^{
\resizebox{0.2in}{!}
{\begin{ytableau}
       ~ \\
       ~ \\
       ~ \\
       ~ \\
       ~ & ~\\
\end{ytableau}
}}
=
\xi_1\mp (2\epsilon_+-6\epsilon_-),
\quad
\xi_{1,\pm}^{
\resizebox{0.1in}{!}
{\begin{ytableau}
       ~ \\
       ~ \\
       ~ \\
       ~ \\
       ~ \\
       ~ \\
\end{ytableau}
}}
=
\xi_1\pm 6(\epsilon_+-\epsilon_-).
$
}
\eea
\be
\phantom{.}
\ee
As in the previous examples, by comparing series expansions we find that equation \eqref{eq:3str} coincides with the expression obtained from equation \eqref{eq:ellreso}.

\section*{Acknowledgments}
We are grateful to Kaiwen Sun and Richard Szabo for helpful discussions.  We would like to thank Institut Mittag-Leffler for hospitality during the workshop ``Enumerative Invariants, Quantum Fields and String Theory Correspondences'' where part of this research was conducted. The work of MDZ and GL has received funding from the European Research Council (ERC) under the Horizon 2020 (grant agreement No. 851931) and Horizon Europe (grant agreement No. 101078365) research and innovation programs. MDZ also acknowledges support from the Simons Foundation (grant \#888984, Simons Collaboration on Global Categorical Symmetries).

\appendix

\section{Comparison with the 5d $\mathcal{N}=1^*$ $U(2)$ partition function on $S^1\times \mathbb{C}^2/\mathbb{Z}_2$}
\label{sec:compar}
In this appendix we compare the rank $r=2$ M-string partition function on $T^2\times\mathbb{C}^2/\mathbb{Z}_2$ with the equivariant partition function of $\mathcal{N}=1^*$ $U(2)$ gauge theory on $S^1\times\mathbb{C}^2/\mathbb{Z}_2$, for which simple expressions can be inferred from results given in the literature for the partition function of 4d $\mathcal{N}=2^*$ $U(2)$ SYM on $\mathbb{C}^2/\mathbb{Z}_n$ \cite{Bruzzo:2013daa}. The two expressions do not match on the nose, but we find that by making a few specific modifications to the gauge theory partition function the two expressions can be made to agree. While this points to subtleties in the definition of the partition function, which in fact have already been noted in \cite{Bruzzo:2013daa}, we view the fact that we are able to match the M-string partition function the gauge theoretic expansion as highly nontrivial evidence for our results and more generally for the duality between the 6d M-string SCFTs and 5d $\mathcal{N}=1^*$ SYM.\\

The $S^1 \times \mathbb{C}^2/\mathbb{Z}_2$ partition function depends on a pair of integers $\boldsymbol{\nu}=(\nu^{(1)},\nu^{(2)}),$ $0\leq \nu^{(a)}< 2$ which label the monodromy of the $U(2)$ gauge field at infinity. By an action of the Weyl group of $U(2)$ one can further restrict to $\nu^{(1)}\leq \nu^{(2)}$.  The partition function takes the following factorized form:
\begin{align}
Z^{\boldsymbol{\nu}}_{U(2),\, S^1\times\mathbb{C}^2/\mathbb{Z}_2}(\epsilon_+,\epsilon_-,\tau,\boldsymbol{a},\mu,\xi)
=
Z^{\boldsymbol{\nu}}_{pert}(\epsilon_+,\epsilon_-,\tau,\mu,\xi)
Z^{\boldsymbol{\nu}}_{inst}(\epsilon_+,\epsilon_-,\tau,\boldsymbol{a},\mu,\xi),
\label{eq:zis}
\end{align}
where $\mu$ denotes the mass of the adjoint hypermultiplet, $\tau = \frac{4\pi^2}{g^2}$ is the inverse gauge coupling, and $\boldsymbol{a} = (a^{(1)},a^{(2)})$ are the vevs of the scalars in the vector multiplet. In the 5d Coulomb branch chamber $a^{(2)}> a^{(1)}$, we use  $a^{(2)}-a^{(1)}$ as a parameter for the paritition function. We also define $A^{(a)}=e^{2\pi i a^{(a)}}$.
We employ the following expression for the perturbative partition function:
\begin{align}
\nonumber
Z_{pert}^{\boldsymbol{\nu}}(\epsilon_+,\epsilon_-,\tau,\boldsymbol{a},\mu,\xi)  = &
\prod_{a=1}^2\prod_{\substack{i,j=0\\i+j+1=0\text{ mod } 2}}^\infty\frac{1-e^{2\pi i m}x^{i+j+1}t^{i-j}}{1-x^{i+j+1}t^{i-j+1}}\\
&\times \widetilde Z_{pert}^{\boldsymbol{\nu}}(\epsilon_+,\epsilon_-,\tau,\boldsymbol{a},\mu,\xi),
\label{eq:zperspl}
\end{align}
where the first line of the right hand side of \eqref{eq:zperspl} is the contribution of the massless vector and hypermultiplets, and
\begin{align}
\widetilde Z_{pert}^{\boldsymbol{\nu}}(\epsilon_+,\epsilon_-,\tau,\boldsymbol{a},\mu,\xi)
=
&
\hskip-.3in
\prod_{\substack{i,j=0\\i+j+1=\nu^{(2)}-\nu^{(1)}\text{ mod } 2}}^\infty
\hskip-.3in
\frac{1-\frac{A^{(2)}}{A^{(1)}}e^{2\pi i m}x^{i+j+1}t^{i-j}}{1-\frac{A^{(2)}}{A^{(1)}}x^{i+j+1}t^{i-j+1}}\frac{1-\frac{A^{(2)}}{A^{(1)}}e^{-2\pi i m}x^{i+j+1}t^{i-j}}{1-\frac{A^{(2)}}{A^{(1)}}x^{i+j+1}t^{i-j-1}}.\label{eq:zpertgauge}
\end{align}
are the BPS contributions to the partition function of a $W$-boson and a pair of hypermultiplets of mass $a^{(2)}-a^{(1)}\pm m$. After some manipulations up to overall prefactors that require regularization, this expression can be rewritten as:
\begin{align}
\widetilde Z_{pert}^{\boldsymbol{\nu}}(\epsilon_+,\epsilon_-,\tau,\boldsymbol{a},\mu,\xi)=
&\Gamma^{\mathcal{N}=1^*}_{-\mu}(\frac{A^{(1)}}{A^{(2)}}\kappa_1^{\boldsymbol{\nu}}\left(\frac{t}{x}\right)^{-2};x^{-2},\frac{t^2}{x^2})\nonumber\\
\times&\Gamma^{\mathcal{N}=1^*}_{-\mu}(\frac{A^{(2)}}{A^{(1)}}\kappa_1^{\boldsymbol{\nu}}\left(\frac{t}{x}\right)^{-2};x^{-2},\frac{t^2}{x^2})\nonumber\\
\times&\Gamma^{\mathcal{N}=1^*}_{-\mu}(\frac{A^{(1)}}{A^{(2)}}\kappa_2^{\boldsymbol{\nu}}\left(t x\right)^{-2};x^{2},t^2x^2)\nonumber\\
\times&\Gamma^{\mathcal{N}=1^*}_{-\mu}(\frac{A^{(2)}}{A^{(1)}}\kappa_2^{\boldsymbol{\nu}}\left(t x\right)^{-2};x^{2},t^2x^2),
\label{eq:zpertgauge}
\end{align}
where
\be
\Gamma^{\mathcal{N}=1^*}_\mu(X;t_1,t_2) = \prod_{i=0}^\infty\prod_{j=0}^\infty\frac{1-X \exp(2\pi i\mu) t_1^{-i} t_2^{-j}}{1-X t_1^{-i} t_2^{-j}}
\ee
and we take
\begin{align}
\kappa_1^{\boldsymbol{\nu}} = \left(\frac{t}{x}\right)^{(\nu^{(1)}-\nu^{(2)}) \text{ mod } 2},\qquad \kappa_2^{\boldsymbol{\nu}} = \left(tx \right)^{(\nu^{(1)}-\nu^{(2)}) \text{ mod } 2}.
\label{eq:kappaour}
\end{align}

In the limit of small $S^1$ radius, equation \eqref{eq:zpertgauge} almost reduces to the 4d perturbative partition function on $\mathbb{C}^2/\mathbb{Z}_2$ as given in section 6.3 of \cite{Bruzzo:2013daa}, with a few subtle differences. Namely, to obtain the expressions of \cite{Bruzzo:2013daa} we would need to replace $-\mu$ by $\mu$ in the argument of $\Gamma^{\mathcal{N}=1^*}_{-\mu}$. Also, rather than \eqref{eq:kappaour} we would need to take
\begin{align}
\kappa_1^{\boldsymbol{\nu}} = x^{-2\cdot (\nu^{(1)}-\nu^{(1)} \text{ mod } 2)},\qquad \kappa^{\boldsymbol{\nu}}_2= 1.
\end{align}
However, this choice does not give rise to an expression in terms of BPS particle contributions analogous to \eqref{eq:zpertgauge}, which is natural from a physics perspective. We comment further on this and related subtleties at the end of this section.\\

Let us now move on to the instanton part of the 5d partition function, which we take to be given by:
\begin{align}
Z^{\boldsymbol{\nu}}_{inst}(\epsilon_+,\epsilon_-,\tau,\boldsymbol{a},\mu,\xi)&
=
\hspace{-.35in}
\sum_{\substack{\boldsymbol{v}\in \frac{1}{2}\mathbb{Z}^2\\2v^{(a)}
=
\nu^{(a)} \text{ mod 2 }}}
\hspace{-.35in}
\xi^{v^{(1)}+v^{(2)}}q^{(v^{(1)})^2+(v^{(2)})^2}
\prod_{a,b=1}^2\frac{\ell_{v^{(b)}-v^{(a)}}(t^2 x^2,x^{-2},\frac{A^{(b)}}{A^{(a)}}e^{-2\pi i \mu})}{\ell_{v^{(b)}-v^{(a)}}(t^2 x^2,x^{-2},\frac{A^{(b)}}{A^{(a)}})}\nonumber\\
&
\hspace{-.25in}
\times
Z^{inst}_{\mathbb{C}^2}(\epsilon_+,\epsilon_++2\epsilon_-,\boldsymbol{a}+2(\epsilon_++\epsilon_-) \boldsymbol{v},\mu,q)\nonumber\\
&
\hspace{-.25in}
\times
Z^{inst}_{\mathbb{C}^2}(\epsilon_+,-\epsilon_++2\epsilon_-,\boldsymbol{a}+2(\epsilon_+-\epsilon_-) \boldsymbol{v},\mu,q),
\label{eq:zinstU2}
\end{align}
where we take the following definitions for the \emph{edge contributions}:
\begin{align}
\label{eq:edge}
\ell_v(t_1,t_2;X)
=
\begin{cases}
\prod_{i=0}^{\lfloor{v}\rfloor-1}\prod_{j=0}^{2i+2(v \text{ mod } 1)}\frac{1-X t_1^{i+(v\text{ mod }1)} t_2^j}{(X t_1^{i+(v\text{ mod }1)} t_2^j)^{1/2}}\qquad \hspace{.25in} \lfloor{v}\rfloor>0\\
1 \qquad \hspace{2.74in} \lfloor{v}\rfloor=0\\
\prod_{i=0}^{-\lfloor{v}\rfloor}\prod_{j=0}^{2i-2(v \text{ mod } 1)-1}\frac{1-X t_1^{-i+(v\text{ mod }1)} t_2^{-j}}{(X t_1^{-i+(v\text{ mod }1)} t_2^{-j})^{1/2}}\qquad  \lfloor{v}\rfloor<0
\end{cases}
\end{align}
while for the instanton partition function on $S^1\times \mathbb{C}^2$ we take \cite{Haghighat:2013gba}:
\begin{align}
Z^{inst}_{\mathbb{C}^2}(\epsilon_+,\epsilon_-,\tau,\boldsymbol{a},\mu)
\hskip-.05in
=
\hskip-.05in
\sum_{\vec Y} \left(\frac{q}{e^{4\pi i \mu}}\right)^{\vert Y_1\vert+\vert Y_2\vert}
\hskip-.05in
\prod_{a,b=1}^2
&\prod_{(i,j)\in Y_a}
\hskip-.1in
\frac{1-e^{2\pi i \mu}\frac{A_b}{A_a}(tx)^{-Y^t_{b,j}+i} (t/x)^{-Y_{a,i}-j+1}}{1-\frac{A_b}{A_a}(tx)^{-Y^t_{b,j}+i} (t/x)^{-Y_{a,i}-j+1}}
\nonumber\\
\times
&\hskip-.1in
\prod_{(i,j)\in Y_b}\frac{1-e^{2\pi i \mu}\frac{A_b}{A_a}(tx)^{Y^t_{a,j}-i+1} (t/x)^{-Y_{b,i}+j}}{1-\frac{A_b}{A_a}(tx)^{Y^t_{a,j}-i+1} (t/x)^{-Y_{b,i}+j}}.
\end{align}
Here, $\vec Y = (Y_1,Y_2)$ denotes a pair of Young diagrams. As for the perturbative factors, these expressions do not quite reduce in the 4d limit to the ones of \cite{Bruzzo:2013daa}, but can be made to coincide upon making certain small adjustments to the expressions in \cite{Bruzzo:2013daa}. Namely, in the first line of equation \eqref{eq:zinstU2} one needs to replace $\mu \to -\mu$ in the argument of the {edge contributions}. Moreover, for the 4d leg factors \cite{Bruzzo:2013daa} take a slightly different dependence on $(v\text{ mod }1)$, which in 5d would correspond to:
\begin{align}
\label{eq:edge}
\ell_v(t_1,t_2;X)
=
\begin{cases}
\prod_{i=0}^{\lfloor{v}\rfloor-1}\prod_{j=0}^{2i+2(v \text{ mod } 1)}\frac{1-X t_1^{i} t_2^j}{(X t_1^{i} t_2^j)^{1/2}}\qquad \hspace{.77in} \lfloor{v}\rfloor>0\\
1 \qquad \hspace{2.74in} \lfloor{v}\rfloor=0\\
\prod_{i=0}^{-\lfloor{v}\rfloor}\prod_{j=0}^{2i-2(v \text{ mod } 1)-1}\frac{1-X t_1^{-i+2(v\text{ mod }1)} t_2^{-j}}{(X t_1^{-i+(v\text{ mod }1)} t_2^{-j})^{1/2}}\qquad  \lfloor{v}\rfloor<0
\end{cases}
\end{align}
Indeed, the authors of \cite{Bruzzo:2013daa} stress in Remark (C.12) that the expressions for the edge contributions, as well as for the perturbative contributions, depend on a discrete choice: we attribute the slight mismatch we find to this fact. It would be interesting to see if it is possible to reproduce our result by making an appropriate choice within the framework of \cite{Bruzzo:2013daa}.\footnote{We would like to thank R. Szabo for a discussion of this point.} 

We conjecture that the partition function of $U(2)$ $\mathcal{N}=1^*$ SYM on $S^1\times\mathbb{C}^2/\mathbb{Z}_2$, given by equation \eqref{eq:zis} with the choices mentioned in the previous paragraphs, coincides with the M-string partition function on $T^2\times\mathbb{C}^2/\mathbb{Z}_2$, which explicitly is given by:
\bea
\nonumber
&&\hskip-0.75in
\mathcal{Z}^{\boldsymbol{\omega}^{KK}}_{T^2\times\mathbb{C}^2/\mathbb{Z}_2}(\vec{\xi},\epsilon_+,\epsilon_-,m,\tau)
=
q^{\frac{2n}{24}}
\times
\mathcal{Z}^{\text{BPS particles}}_{T^2\times\mathbb{C}^2/\mathbb{Z}_2}(\epsilon_+,\epsilon_-,m,\tau)
\\
&\times&
\sum_{\kappa=0}^\infty
e^{-\kappa \varphi} \mathbb{E}^{\boldsymbol{\omega}^{KK}}_{\kappa}(\vec{\xi},\epsilon_+,\epsilon_-,m,\tau),
\eea
where for $\Im a^{(2)}> \Im a^{(1)}$ we make the identifications
\be
\omega^{KK,(a)} = \nu^{KK,(a)},
\qquad
e^{-\varphi} = -\frac{A^{(2)}}{A^{(1)}},
\ee
while for $\Im a^{(2)}<\Im a^{(1)}$ we make the identification
\be
\omega^{KK,(a)} = \nu^{KK,(3-a)},
\qquad
e^{-\varphi} = -\frac{A^{(1)}}{A^{(2)}}.
\ee

We have verified this conjecture by expanding the gauge theoretic and M-string partition function up to string number $\kappa = 4$ and instanton number $4$, finding exact agreement.  Here we report the first few coefficients in the series expansion of the partition function for inequivalent choices of monodromy. For trivial monodromy we find:
\bea
\nonumber
&&
\mathcal{Z}^{(0,0)}_{T^2\times\mathbb{C}^2/\mathbb{Z}_2}(\vec{\xi},\epsilon_+,\epsilon_-,m,\tau) =
\left[1+ 2q\left(
N \mathcal{H}
+
\left(\xi^{-\frac{1}{2}}+\xi^{\frac{1}{2}}\right)^2
\right)+\mathcal{O}(q^2)\right]
\nonumber
\\
\nonumber
&&
-e^{-\varphi}N\bigg[\mathcal{H}+q
\left(
2\mathcal{H}
-
t^{-1}-t
\right)
\left(
N\mathcal{H}+
\left(\xi^{-\frac{1}{2}}+\xi^{\frac{1}{2}}\right)^2
\right)
+\mathcal{O}(q^2)
\bigg]
+
\mathcal{O}\left(e^{-2\varphi}\right),
\eea
where
\be
N = \frac{(1-Q_m t)(1-Q_m t^{-1})}{Q_m}\quad\text{and}\quad\mathcal{H}=\mathcal{H}_{\mathbb{C}^2/\mathbb{Z}_2}=\frac{t(1+t^2)}{(1-t^2 x^2)(1-t^2x^{-2})}.
\ee
For monodromy $\boldsymbol{\omega}^{KK}=(1,0)$ or $(0,1)$ we find: 
\bea
\nonumber
&&
\mathcal{Z}^{(1,0)}_{T^2\times\mathbb{C}^2/\mathbb{Z}_2}(\vec{\xi},\epsilon_+,\epsilon_-,m,\tau)=\mathcal{Z}^{(0,1)}_{T^2\times\mathbb{C}^2/\mathbb{Z}_2}(\vec{\xi},\epsilon_+,\epsilon_-,m,\tau) =
\\
\nonumber
&&
\left(\xi^{-\frac{1}{2}}+\xi^{\frac{1}{2}}\right)q^{\frac{1}{4}}
\bigg\{\left[1+ q\left(\xi^{-1}+4+\xi+2
N \mathcal{H}\right)
+\mathcal{O}(q^2)\right]
\nonumber
\\
\nonumber
&&
-e^{-\varphi}N\mathcal{H}\frac{x+\frac{1}{x}}{t+\frac{1}{t}}
\bigg[1+q
\bigg(
\xi^{-1}+4+\xi+2N \mathcal{H}
\!+\!
\left(x^{-2}+1+x^2\right)\!\left(t^{-2}+1+t^2\right)
\\
\nonumber
&&
\!-\!
\left(t^{-4}+t^{-2}+1+t^2+t^4\right)
\!-\!
\left(t^{-1}+t\right)\!\left(Q_m^{-1}+Q_m\right)
\!
\bigg)
+\mathcal{O}(q^2)
\bigg]
+
\mathcal{O}\left(e^{-2\varphi}\right)\bigg\}.
\eea
Finally, for monodromy $\boldsymbol{\omega}^{KK}=(1,1)$ we obtain: 
\bea
\nonumber
&&
\mathcal{Z}^{(1,1)}_{T^2\times\mathbb{C}^2/\mathbb{Z}_2}(\vec{\xi},\epsilon_+,\epsilon_-,m,\tau) =
\\
\nonumber
&&
q^{\frac{1}{2}}
\bigg\{
\left(\xi^{-\frac{1}{2}}+\xi^{\frac{1}{2}}\right)^2
\left[1+ q\left(4+2
N \mathcal{H}\right)
+\mathcal{O}(q^2)\right]
\nonumber
\\
\nonumber
&&
e^{-\varphi}N
\bigg[
\mathcal{H}\left(\xi^{-\frac{1}{2}}+\xi^{\frac{1}{2}}\right)^2-t^{-1}-t
+
q
\bigg(
-\left(t^{-1}+t\right)\left(t^{-2}+t^2\right)\left(x^{-2}+x^2\right)\\
\nonumber
&&+\left((N\mathcal{H}+2)\left(\xi^{-\frac{1}{2}}+\xi^{\frac{1}{2}}\right)^2-N(t^{-1}+t)\right)\left(2\mathcal{H}-t^{-1}-t\right)
\bigg)
+\mathcal{O}(q^2)
\bigg]
+
\mathcal{O}\left(e^{-2\varphi}\right)\bigg\}.
\eea

\bibliography{untitled}
\bibliographystyle{utphys}

\end{document}